\documentstyle[aps,psfig,prd,preprint,tighten]{revtex}

\begin{document}
\vspace{1.0cm}

\title{Nucleon Charge and Magnetization Densities from Sachs Form Factors}
\author{James J. Kelly}
\address{ Department of Physics, University of Maryland, 
          College Park, MD 20742 }
\date{September 27, 2002}
\maketitle

\begin{abstract}
Relativistic prescriptions relating Sachs form factors to nucleon
charge and magnetization densities are used to fit recent data for
both the proton and the neutron.
The analysis uses expansions in complete radial bases to minimize
model dependence and to estimate the uncertainties in radial densities
due to limitation of the range of momentum transfer.
We find that the proton charge distribution, fitted to recent 
recoil-polarization data displaying an almost linear decrease in
$G_{Ep}/G_{Mp}$ for $Q^2 \gtrsim 1$ (GeV/$c$)$^2$, is significantly
broader than its magnetization density.
We also find that the magnetization density is broader for the neutron 
than the proton.
The neutron charge form factor is consistent with the Galster parametrization
over the available range of $Q^2$, but the relativistic inversion produces
a softer radial density.
Discrete ambiguities in the inversion method are analyzed in detail.
The method of Mitra and Kumari ensures compatibility with pQCD at
large $Q^2$ and is most useful for extrapolating form factors.
Although a recent observation that $Q F_{2p}/F_{1p}$ is approximately
constant for $2<Q^2<6$ (GeV/$c$)$^2$ appears to be inconsistent
with the $Q^{-2}$ scaling expected from quark helicity conservation, 
our analysis fits these data while remaining consistent with pQCD for 
large $Q^2$.
\end{abstract}

\section{Introduction}
\label{sec:intro}

A fundamental test of the QCD confinement mechanism, 
either on the lattice or in models,
is the electromagnetic structure of the nucleon.
This electromagnetic structure is reflected by the electric
and magnetic form factors, $G_E(Q^2)$ and $G_M(Q^2)$, measured 
by electron scattering with spacelike invariant momentum
transfer $Q$.
At low $Q$ one can interpret these form factors as Fourier
transforms of the nucleon charge and magnetization densities
\cite{Ernst60,Sachs62}, 
but the relationship between form factor and density is
complicated by recoil as $Q$ increases.
Although models of nucleon structure can often calculate
the form factor directly, it is still desirable to relate form 
factors to spatial densities because our intuition tends to be 
grounded more firmly in space than momentum transfer.
In this paper we fit charge and magnetization densities to 
recent nucleon form factor data using a prescription that
accounts for nucleon recoil and Lorentz contraction and is
compatible with perturbative QCD (pQCD) scaling at large $Q^2$.

Early experiments with modest $Q^2$ suggested that 
\begin{equation}
\label{eq:dipole}
G_{Ep} \approx \frac{G_{Mp}}{\mu_p} \approx \frac{G_{Mn}}{\mu_n}
\approx G_D
\end{equation}
where $G_D(Q^2) = (1 + Q^2/\Lambda^2)^{-2}$ with 
$\Lambda^2 = 0.71$ (GeV/$c$)$^2$ is known as the dipole form factor 
\cite{Hughes65,Dunning66}.
However, the naive Fourier transform of the dipole form factor produces
an exponential density with an unphysical cusp at the origin.
Similarly, data for $G_{En}$ at low $Q^2$ can be described by the 
Galster parametrization \cite{Galster71}
\begin{equation}
\label{eq:Galster}
G_{En}(Q^2) \approx -\mu_n G_D(Q^2) \frac{A \tau}{1+B\tau}
\end{equation}
where $A$ and $B$ are constants and $\tau=(Q/2m)^2$,
but direct Fourier transform of this form factor also produces
a cusp at the origin.
Using a relativistic inversion formula that accounts for the Lorentz
contraction along the momentum transfer,
Licht and Pagnamenta \cite{Licht70a} obtained a reasonable fit
to proton form factors using a Gaussian density with a more realistic
behavior in the interior.
Ji \cite{Ji91} obtained similar results using a relativistic soliton
model and we used this model in Ref.\ \cite{Kelly01} to fit data
for Sachs form factors.
These models offer plausible radial densities, but are not compatible
with pQCD scaling unless one imposes somewhat awkward restrictions upon 
the Fourier transform, as discussed in Sec.\ \ref{sec:pQCD}.
Fortunately, a variant proposed by  Mitra and Kumari \cite{Mitra77} 
complies with pQCD scaling automatically, without need of such constraints.
We use this method, described as {\it relativistic inversion}, to extract 
nucleon charge and magnetization densities from data for Sachs form factors.
Our fitting procedure minimizes model dependence by employing linear 
expansions in complete radial bases, 
such as Fourier-Bessel or Laguerre-Gaussian expansions,
and estimates uncertainties arising from the limitation of experimental
data to a finite range of momentum transfer using methods originally 
developed to analyze electron scattering by nuclei.
Such an analysis produces a good fit to form factor data using a radial
density whose error band reflects both the statistical quality of the
data and its limited coverage of momentum transfer.
Differences between densities obtained using several variations of the
inversion formula are described as {\it discrete ambiguities} and are 
analyzed in detail herein.

Data for $G_{Mp}$ and $G_{Mn}$ with $Q^2>1$ (GeV/$c$)$^2$ show significant
departures from the simple dipole parametrization, 
but the extraction of $G_{Ep}$ using the traditional Rosenbluth method 
\cite{Rosenbluth50}
becomes increasingly difficult as $Q^2$ increases because the dominance of 
the $G_{Mp}$ contribution to the cross section increases with $Q^2$.
Consequently, there are large statistical uncertainties in Rosenbluth
data for $G_{Ep}$ at $Q^2 > 1$ (GeV/$c$)$^2$ and the discrepancies between 
comparable experiments suggests that systematic errors in the Rosenbluth
analysis are often underestimated \cite{Kelly96}.
More recently, recoil polarization has been used to measure the ratio
$g_p = G_{Ep}/G_{Mp}$ directly, without need of Rosenbluth separation.
In this technique, the components of the nucleon polarization 
$\vec{P}^\prime$ after scattering by a polarized electron beam are measured 
along the momentum transfer direction, denoted by $\hat{z}$, and in
the $\hat{x}$ direction transverse to $\hat{z}$ in the scattering plane.  
The form factor ratio is then obtained using \cite{Dombey69a,Arnold81}
\begin{equation}
\label{eq:g}
\frac{P^\prime_x}{P^\prime_z} = -\sqrt{\frac{2\epsilon}{\tau (1+\epsilon)}} 
\; g
\end{equation}
where $\epsilon = (1+(1+\tau)2 \tan^2{\theta_e/2})^{-1}$ is
the transverse polarization of the virtual photon for an electron scattering
angle $\theta_e$.
For the proton, both components can be measured simultaneously using
a polarimeter in the focal plane of a magnetic spectrometer, thereby
minimizing systematic uncertainties due to beam polarization, analyzing power,
and kinematic parameters.
The systematic uncertainty due to precession of the proton spin in the
magnetic spectrometer is usually much smaller than the systematic uncertainties
in combining the absolute cross sections obtained with different kinematical 
conditions and acceptances that are needed for the Rosenbluth method.
Recent data using the recoil polarization technique 
\cite{MKJones00,Gayou01,Gayou02}
have shown a dramatic, almost linear, 
decrease in $G_{Ep}/G_{Mp}$ for $Q^2 > 1$ (GeV/$c$)$^2$.
It was suggested that those results demonstrate that the proton charge is
distributed over a larger volume than its magnetization, 
but radial densities were not obtained.  
Our analysis confirms that interpretation quantitatively. 

Similar techniques can be used to obtain the neutron form factor ratio,
$g_n=G_{En}/G_{Mn}$, using either target or recoil polarization,
but in the absence of a target with free neutrons one must employ 
quasifree scattering from a neutron bound in a light nucleus.
Detection of a recoil neutron with quasifree kinematics and small
missing momentum tends to minimize uncertainties due to nuclear structure
and final state interactions \cite{Arenhovel87}.
Although considerable care is still needed at low $Q^2$, polarization
methods offer smaller systematic errors and less model dependence than 
traditional Rosenbluth analyses of elastic scattering or quasifree knockout.
We extracted the neutron charge density from recent polarization data
for $g_n$ for $Q^2 < 1.6$ (GeV/$c$)$^2$ using the relativistic inversion
method.
Although the form factor data remain consistent with the Galster
parametrization over this range of momentum transfer, 
the charge density obtained by relativistic inversion is considerably
softer than that from nonrelativistic inversion of the Galster form
factor and does not feature a cusp at the origin.
Over the next several years, extending of the experimental range of 
momentum transfer should substantially reduce the uncertainty in the
interior density.

The model is presented in Sec.\ \ref{sec:model}, the analysis procedures
in Sec.\ \ref{sec:procedure}, and principal results in Sec.\ \ref{sec:results}.
In Sec.\ \ref{sec:discussion} we compare our results to another analyses,
discuss the extrapolation to higher $Q^2$ and the role of 
discrete ambiguities in fitted densities.
We also form combinations of neutron and proton charge densities that in 
the naive quark model represent the distribution of up and down quarks
in the proton.
Finally, our conclusions are summarized in Sec.\ \ref{sec:conclusions}.

\section{Model}
\label{sec:model}

\subsection{Sachs Form Factors}
\label{sec:sachs}

Matrix elements of the nucleon electromagnetic current operator $J^\mu$
take the form
\begin{equation}
\langle N(p^\prime,s^\prime) | J^\mu | N(p,s) \rangle =
\bar{u}(p^\prime,s^\prime) e \Gamma^\mu u(p,s)
\end{equation}
where $u$ is a Dirac spinor, $p,p^\prime$ are initial and final momenta,
$q=p-p^\prime$ is the momentum transfer, $s,s^\prime$ are spin four-vectors, 
and where the vertex function
\begin{equation}
\Gamma^\mu = F_1(Q^2)\gamma^\mu + 
\kappa F_2(Q^2) \frac{i \sigma^{\mu\nu}q_\nu}{2m}
\end{equation}
features Dirac and Pauli form factors $F_1$ and $F_2$ that depend upon
the nucleon structure.
Here $e$ is the elementary charge, $m$ is the nucleon mass, 
$\kappa$ is the anomalous part of the magnetic moment, and 
$\gamma^\mu$ and $\sigma^{\mu\nu}$ are the usual Dirac matrices 
({\it e.g.}, \cite{BjorkenDrella}).
The interpretation of these form factors appears simplest in the nucleon
Breit frame where the energy transfer vanishes.
In this frame the nucleon approaches with initial momentum
$-\vec{q}_{B}/2$, receives three-momentum transfer $\vec{q}_B$, 
and leaves with final momentum $\vec{q}_B/2$.
Thus, the nucleon Breit frame momentum is defined by
$q_B^2 = Q^2 = q^2 / (1+\tau)$ where $(\omega,\vec{q})$ 
is the momentum transfer in the laboratory, 
$Q^2 = q^2 - \omega^2$ is the spacelike invariant four-momentum 
transfer, and $\tau=Q^2/4m^2$.
In the Breit frame for a particular value of $Q^2$, the current  
separates into electric and magnetic contributions \cite{Sachs62}
\begin{equation}
\label{eq:BreitCurrent}
\bar{u}(p^\prime,s^\prime) \Gamma^\mu u(p,s) = 
\chi^\dagger_{s^\prime}
\left( G_E + \frac{i\vec{\sigma}\times\vec{q}_B}{2m} G_M 
\right) \chi_s
\end{equation}
where $\chi_s$ is a two-component Pauli spinor and where the
Sachs form factors are given by
\begin{mathletters}
\begin{eqnarray}
G_E &=& F_1 - \tau \kappa F_2 \\
G_M &=& F_1 + \kappa F_2
\end{eqnarray}
\end{mathletters}
The similarity of Eq. (\ref{eq:BreitCurrent}) to the classical current density
\begin{equation}
J^{NR} = \left( 
e \rho^{NR}_{ch}, \mu \vec{\sigma}\times \vec{\nabla} \rho^{NR}_m \right)
\end{equation}
suggests an identification of charge and magnetization densities
\begin{mathletters}
\label{eq:NR}
\begin{eqnarray}  
\rho^{NR}_{ch}(r) &=& \frac{2}{\pi} \int_0^\infty dQ \; Q^2 j_0(Qr) G_E(Q^2) \\
\mu \rho^{NR}_{m}(r) &=& \frac{2}{\pi} 
\int_0^\infty dQ \; Q^2 j_0(Qr) G_M(Q^2)
\end{eqnarray}
\end{mathletters}
where $\mu=1+\kappa$ is the appropriate static magnetic moment 
(either $\mu_p$ or $\mu_n$) relative to the nuclear magneton.
However, this naive inversion procedure is described as nonrelativistic (NR)
because it ignores the variation of the Breit frame with $Q^2$.

\subsection{Intrinsic Form Factors}
\label{sec:intrinsic}

Let $\rho_{ch}(r)$ and $\rho_{m}(r)$ represent spherical 
charge and magnetization densities in the nucleon rest frame.
It is convenient to normalize these intrinsic densities according to
\begin{mathletters}
\begin{eqnarray}
\int_0^\infty dr \; r^2 \rho_{ch}(r) &=& Z \\
\int_0^\infty dr \; r^2 \rho_{m}(r) &=& 1
\end{eqnarray}
\end{mathletters}
where $Z=0,1$ is the nucleon charge.
Fourier-Bessel transforms of the intrinsic densities are defined by
\begin{equation}
\tilde{\rho}(k) = \int_0^\infty dr \; r^2 j_0(kr) \rho(r)
\end{equation}
where $k$ is the spatial frequency (or wave number).
We describe $\tilde{\rho}(k)$ as an {\it intrinsic form factor}.
If one knew how to obtain $\tilde{\rho}(k)$ from data for the
appropriate Sachs form factor, the intrinsic density could be
obtained simply by inverting the Fourier transform, such that
\begin{equation}
\label{eq:rhor}
\rho(r) = \frac{2}{\pi} \int_0^\infty dk \; k^2 j_0(kr) \tilde{\rho}(k)
\end{equation}

The naive nonrelativistic inversion method assumes that 
$k \rightarrow Q$ and $\tilde{\rho}(Q) \rightarrow G(Q^2)$
where $G(Q^2)$ is the appropriate Sachs form factor.
However, this inversion procedure produces unsatisfactory results
for the common dipole and Galster parametrizations ---
the corresponding radial densities have unphysical cusps at the
origin and rather hard cores.
For example, the naive Fourier transform of the dipole form factor 
produces an exponential density.
(Although it appears much more complicated, the Galster density 
can also be obtained in closed form and displays similarly
unrealistic behavior near the origin.)
Licht and Pagnamenta \cite{Licht70b} attributed these failures of
nonrelativistic inversion to the replacement of the intrinsic
spatial frequency $k$ with the momentum transfer $Q$ and demonstrated 
that by applying a boost from the Breit frame with momentum $q_B = Q$ 
to the rest frame, inversion of the dipole form factor using a reduced 
spatial frequency
\begin{equation}
\label{eq:k}
k^2 = \frac{Q^2}{1+\tau}
\end{equation}
softens the density.
In fact, a good fit to the data for $G_{Ep}$ could then be obtained
using a Gaussian density typical of quark models.

Unfortunately, unique relativistic relationships between the Sachs form 
factors measured by electron scattering at finite $Q^2$ and the static 
charge and magnetization densities in the nucleon rest frame do not exist.
The basic problem is that electron scattering measures transition matrix
elements between states of a composite system that have different momenta
and the transition densities between such states are different from the 
static densities in the rest frame.
Furthermore, the boost operator for a composite system depends upon the 
interactions among its constituents.
Nevertheless, a wide variety of models have employed similar relativistic
prescriptions to relate elastic form factors to ground-state densities.
The first proposal was made by Licht and Pagnamenta \cite{Licht70b} 
using a cluster model and a kinematic boost that neglects interactions.
The transition form factors were evaluated using the impulse 
approximation and neglecting relative motion.
Mitra and Kumari \cite{Mitra77} obtained similar results using a 
kinematic transformation that is more symmetric between initial and
final states and can be applied to inelastic scattering also. 
Ji \cite{Ji91} also obtained similar results using a relativistic Skyrmion 
model based upon a Lorentz invariant Lagrangian density for which the
classical soliton solution can be evaluated in any frame.
Quantum fluctuations were then evaluated after the boost.
Although an approximation is still required to evaluate the transition form
factors, it was argued that this approximation is best in the Breit frame.
Holzwarth \cite{Holzwarth96} extended the soliton model to the timelike
regime and analyzed the superconvergence relations needed to obtain
spectral functions.

Each of these prescriptions can be represented in the form
\begin{mathletters}
\label{eq:rhok}
\begin{eqnarray}
\tilde{\rho}_{ch}(k) &=& G_E(Q^2) (1+\tau)^{\lambda_E} \\
\mu \tilde{\rho}_{m}(k) &=& G_M(Q^2) (1+\tau)^{\lambda_M} 
\end{eqnarray}
\end{mathletters}
where $G(Q^2)$ is one of the four Sachs form factors,
$k$ is the intrinsic spatial frequency given by Eq.\ (\ref{eq:k}), and
$\lambda$ is a model-dependent constant. 
The most important relativistic effect is Lorentz contraction of  
spatial distributions in the Breit frame and the corresponding increase
of spatial frequency represented by the 
factor of $(1+\tau)$ in Eq. (\ref{eq:k}).
A measurement with Breit-frame momentum transfer $q_B = Q$ probes a 
reduced spatial frequency $k$ in the rest frame.
The Sachs form factor for a large invariant momentum transfer $Q^2$ is 
determined by a much smaller spatial frequency $k^2=Q^2/(1+\tau)$ and 
thus declines much less rapidly with respect to $Q^2$ than the Fourier 
transform of the density declines with respect to $k^2$.
In fact, the accessible spatial frequency is limited to $k \leq 2m$ 
such that the asymptotic Sachs form factors in the limit 
$Q^2 \rightarrow \infty$ are determined by the Fourier transform of 
intrinsic densities in the immediate vicinity of the 
limiting frequency $k_m=2m$.
In this model, no information can be obtained beyond the limiting 
frequency determined by the nucleon Compton wavelength.
This limitation can be understood as a consequence of relativistic 
position fluctuations, known as of {\it zitterbewegung}, that smooth out 
radial variations on scales smaller than the Compton wavelength.

Ji \cite{Ji91} derived $\lambda_E=0$ for electric and
$\lambda_M=1$ for magnetic form factors in the soliton model
and attributed the difference between $\lambda_E$ and $\lambda_M$ to the 
Lorentz transformation properties of scalar and vector densities.
The same choices were employed by Holzwarth \cite{Holzwarth96,Holzwarth02}.
On the other hand, Licht and Pagnamenta \cite{Licht70b} obtained 
$\lambda_E=\lambda_M=1$ using the cluster model, 
but Mitra and Kumari \cite{Mitra77} found that a more symmetric 
treatment of the kinematics gives values
$\lambda_E=\lambda_M=2$ that automatically satisfy the perturbative
QCD scaling relations at very large $Q^2$.
For most of the present analysis we will employ the symmetric choice
$\lambda_E=\lambda_M=2$ because fewer restrictions upon the behavior
of $\tilde{\rho}(k)$ are needed near $k_m$ to ensure compatibility
with the asymptotic behavior of Sachs form factors expected from
dimensional scaling.
We describe the variation of a fitted density with the choice of
$\lambda$ as a {\it discrete ambiguity}.
The effect of discrete ambiguities upon fitted densities will be
examined in Sec.\ \ref{sec:ambiguities}.

\subsection{Asymptotic Behavior}
\label{sec:pQCD}

The present form factor model with $\lambda \geq 0$ suggests that the 
asymptotic behavior for $Q \gg 2m$ is given by
\begin{eqnarray}
G(Q^2) &\simeq&
\left( \frac{k_m}{Q} \right)^{2\lambda} 
\left( \tilde{\rho}(k_m)  -
(2\lambda \tilde{\rho}(k_m) + k_m \tilde{\rho}^\prime(k_m)) \frac{k_m^2}{2Q^2} 
\right.  \nonumber \\
&+& \left.
(4\lambda(\lambda+1)\tilde{\rho}(k_m) + 
(3+4\lambda)k_m \tilde{\rho}^\prime(k_m)
+ k_m^2 \tilde{\rho}^{\prime\prime}(k_m)) \frac{k_m^4}{8Q^4} + ... \right)
\end{eqnarray}
where $\tilde{\rho}^\prime$ and $\tilde{\rho}^{\prime\prime}$ are 
derivatives of the momentum-space density evaluated at the limiting 
frequency $k_m=2m$.
Evidently, noninteger values of $\lambda$ are incompatible with the
perturbative QCD prediction \cite{Brodsky73,Brodsky75}
that $G \simeq Q^{-4}$ aside from logarithmic corrections.
Similarly, $\lambda>2$ is excluded also, leaving just three choices.
If we choose $\lambda=2$, then we need only require $\tilde{\rho}(k_m) \neq 0$
to obtain consistency with pQCD.
Thus, the proposal by Mitra and Kumari \cite{Mitra77} of
$\lambda_E=\lambda_M=2$ offers the most natural approach to the pQCD limit.
If we choose $\lambda=1$, 
as recommended by Refs.\ \cite{Ji91,Holzwarth96}
for $G_M$ or by Ref.\ \cite{Licht70b} for both $G_E$ and $G_M$, 
then we must require $\tilde{\rho}(k)$ to have a node at $k_m$ such that 
$\tilde{\rho}(k_m)=0$ and $\tilde{\rho}^\prime(k_m) \neq 0$.
Finally, if we choose $\lambda=0$ as recommended by 
Refs.\ \cite{Ji91,Holzwarth96} for $G_E$, 
then we must impose the somewhat unnatural constraints
$\tilde{\rho}(k_m)=\tilde{\rho}^\prime(k_m)=0$
with $\tilde{\rho}^{\prime\prime}(k_m) \neq 0$.
Thus, it appears that the usefulness of the chiral soliton model
is limited to $Q^2 \ll 4 m^2$ and in order to fit data for larger $Q^2$
with that model Holzwarth found it necessary to artificially increase
the soliton mass \cite{Holzwarth02}. 

Although the intrinsic form factors $\tilde{\rho}(k)$ obtained using 
either dipole or Galster functions for $G(Q^2)$ in Eq.\ (\ref{eq:rhok})
are compatible with the pQCD constraints upon $\tilde{\rho}(k_m)$,
neither can be inverted using Eq.\ (\ref{eq:rhor}) with $\lambda=0$
because $\tilde{\rho}(k) \simeq G(-k_m^2) (-k_m/k)^{2\lambda}$
for $k \rightarrow \infty$.
The inversion integrals for these functions converge well for $\lambda=2$,
slowly for $\lambda=1$, and diverge for $\lambda=0$.
Recognizing that pQCD favors $\lambda=2$, we expect $\tilde{\rho}(k)$ to
have an asymptotic $k^{-4}$ behavior with an amplitude determined
by the nucleon-antinucleon annihilation process
$N \bar{N} \rightarrow e^-e^+$ at threshold.
Similarly, the behavior of $\tilde{\rho}(k)$ for $k \rightarrow \infty$ 
should be determined by the electromagnetic annihilation data for  
$Q^2 \lesssim -4m^2$.
However, we have not attempted to incorporate electromagnetic data
for timelike $Q^2$ in the present analysis because it is not clear that 
the prescription for intrinsic form factors should apply to that regime.
A more general analysis of the analytic structure of the form factors
can be made using dispersion theory \cite{Hohler76,MMD,HMD},
but that approach does not consider the densities that are the
subject of the present analysis.
Nor do we consider here the modifications of pQCD scaling due
to logarithmic running of the strong coupling \cite{Lepage80}.

With the exception of $G_{Mp}$, the available data constrain 
$\tilde{\rho}(k)$ very little near $k_m$ because the ratio
\begin{equation}
\frac{k}{k_m}  \simeq 1 - \frac{k_m^2}{2Q^2}
\end{equation}
approaches unity relatively slowly as $Q^2$ increases.
Thus, we will find that the choice of $\lambda$ has very little effect
upon the fit to data for Sachs form factors,
but does have a strong effect upon the extrapolation of fitted form factors 
beyond the measured range of $Q^2$.
By incorporating pQCD scaling in its basic parametrization, 
the choice $\lambda=2$ limits the range of variation available to 
extrapolated form factors.
Conversely, without explicit enforcement of pQCD by means of somewhat
artificial constraints upon $\tilde{\rho}(k)$ near $k_m$,
fits with $\lambda<2$ permit much wider latitude at large $Q^2$.
The data for $G_{Mp}$ for $Q^2 > 20$ (GeV/$c$)$^2$ exhibit scaling
and automatically enforce the appropriate constraints upon $\tilde{\rho}(k_m)$,
but data available for the other three electromagnetic form factors do not.
Consequently, one could impose constraints upon $\tilde{\rho}(k_m)$ 
with little effect upon the fits in the measured range of $Q^2$.
However, we chose not to employ constraints of this kind,
which seem rather artificial, and to permit fits with $\lambda<2$ the
greatest possible latitude. 

We use $\lambda_E=\lambda_M=2$ for most of the present work, but will
discuss the consequences of the discrete ambiguity in 
Sec.\ \ref{sec:ambiguities}.
Note that our previous work \cite{Kelly01}, motivated by the soliton model, 
used $\lambda_E=0$ and $\lambda_M=1$.

\subsection{Moments}
\label{sec:moments}

It is customary to describe the low $Q^2$ behavior of a form factor
in terms of a transition radius obtained from integral moments of
the underlying density, but care must be taken with the relativistic
relationship between a Sachs form factor and its intrinsic density. 
We define integral moments by
\begin{equation}
\label{eq:moments1}
M_\alpha = \int_0^\infty dr \; r^{2+\alpha} \rho(r) 
\end{equation}
where $\alpha$ is an even integer.
For a charge density these moments are related to the electric form factor by
\begin{mathletters}
\label{eq:moments}
\begin{eqnarray}
M_0 &=& G(0)  \\ 
M_2 &=& -6 \left. \frac{d G(Q^2) }{dQ^2} \right|_{Q^2\rightarrow 0}
-\frac{3\lambda}{2m^2} G(0)
\end{eqnarray}
\end{mathletters} 
while for magnetization we divide by the magnetic moment.
Thus, one expects $M_0 = Z$ for charge densities and 
$M_0=1$ for magnetization densities.
Notice that the lowest nonvanishing moment is free of discrete ambiguities, 
but that higher moments depend upon $\lambda$.
For example, the mean-square neutron charge radius reduces to
\begin{equation} 
\label{eq:neutron_radius}
\langle r^2 \rangle_n 
= -6 \left. \frac{d G(Q^2) }{dQ^2} \right|_{Q^2\rightarrow 0}
\end{equation}
because the charge vanishes, while the proton radius retains a
small dependence upon $\lambda$
\begin{equation}
\langle r^2 \rangle_{\lambda,p} = \langle r^2 \rangle_{0,p} - 
\frac{3\lambda}{2 m_p^2}
\end{equation}
due to the discrete ambiguity in the intrinsic density.
This term, equal to $0.066\lambda$ in units of fm$^2$, appears to be 
similar to the famous Foldy contribution to the neutron charge radius
\cite{Foldy58}
but has a different origin because it does not depend upon the 
anomalous magnetic moment.

When $M_0 \neq 0$ it is useful to distinguish between a radius
parameter
\begin{equation}
\label{eq:xi}
\xi = \left( -6  \frac{d \ln{G(Q^2)} }{dQ^2} \right)_{Q^2\rightarrow 0}^{1/2}
    = \left( \frac{M_2}{M_0} + \frac{3\lambda}{2m^2} \right)^{1/2}
\end{equation}
based upon the {\it initial logarithmic derivative} of a Sachs form factor 
and the rms radius
\begin{equation}
\label{eq:rms}
\langle r^2 \rangle_\lambda^{1/2} = 
\left( \xi^2 - \frac{3\lambda}{2m^2} \right)^{1/2}
\end{equation}
of the corresponding density obtained for specified $\lambda$.
Thus, $\xi$ is a model-independent property of the
form factor data while $\langle r^2 \rangle^{1/2}$ is
subject to a discrete ambiguity.
These radii agree for $\lambda=0$, but $\langle r^2 \rangle^{1/2}$
is smaller than $\xi$ for larger $\lambda$ due to the zitterbewegung
correction.

Accurate calculations for many phenomena in atomic physics, such as
the Lamb shift, require corrections for the finite size of nucleons.
Although it might appear that the nucleon size should be determined
by an integral moment of a nucleon density through Eq.\ (\ref{eq:moments1}),
the static radial density is not directly measurable by electron scattering.
The discrete ambiguity between the initial slope of the Sachs form
factor and its associated transition density reflects the model-dependence
of the relativistic inversion procedure arising from the treatment of
zitterbewegung.
By convention, QED theorists have decided to identify the radius with
the initial slope of the Sachs form factor and to treat recoil, vacuum
polarization, zitterbewegung, and other effects as separate corrections
({\it e.g.}\ \cite{Friar97,Pachucki95}).
To distinguish between various determinations of nucleon size, we 
describe the model-independent quantity $\xi$ as the {\it Sachs radius}
and the model-dependent rms radius obtained from moments of a fitted radial
density as an {\it intrinsic radius}.
When necessary, these radii are further qualified as either charge
or magnetic and for intrinsic radii by the value of $\lambda$.
The Sachs charge radius is usually the most appropriate for QED applications.

\subsection{Example: Gaussian density}
\label{sec:Gaussian}

\begin{figure}[htbp]
\centerline{ \strut\psfig{file=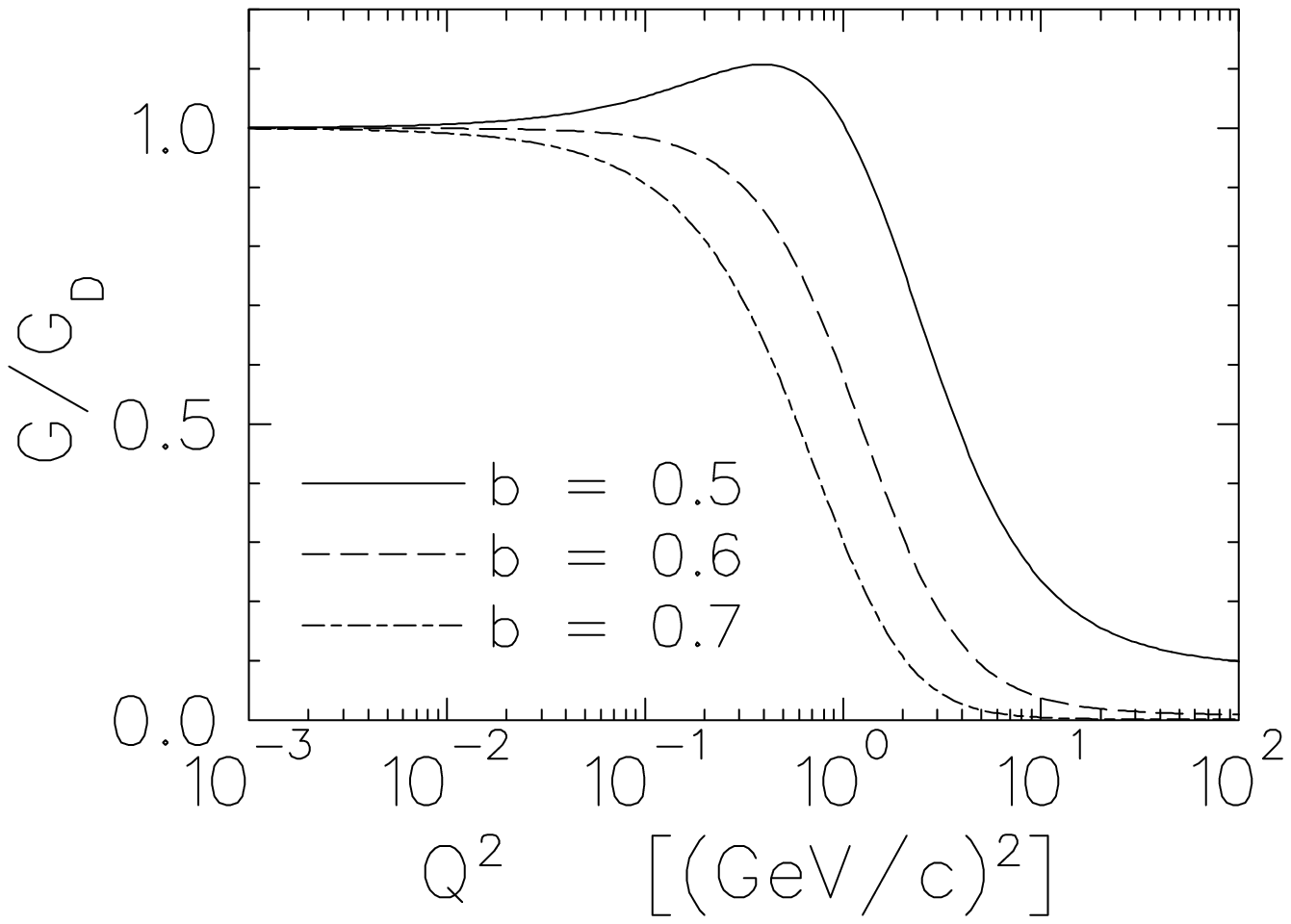,width=4.0in} }
\caption{The ratio between Sachs form factor with $\lambda=2$ and the
dipole form factor is shown for a Gaussian intrinsic density using 
several values of the oscillator parameter, $b$, listed with units in fm.}
\label{fig:Gaussian}
\end{figure}

It is instructive to consider the Sachs form factor that would be produced
by a simple Gaussian density
\begin{equation}
\rho(r) = \frac{4}{b^3\sqrt{\pi}} \exp{(-(r/b)^2)} \hspace{1cm}
\tilde{\rho}(k) = \exp{(-(kb/2)^2)}
\end{equation}
that is typical of quark models. 
The form factor obtained using Eq.\ (\ref{eq:rhok}) with $\lambda=2$
is compared with the familiar dipole form factor in Fig.\ \ref{fig:Gaussian} 
for several choices of $b$; 
note with $b=0.556$ fm the Gaussian parametrization has the same rms 
radius as the dipole form factor.
These curves display the same general features as the data for
$G_{Ep}$, $G_{Mp}$ and $G_{Mn}$:
for low $Q^2$ the form factor is close to the dipole form while for
large $Q^2$ one finds an asymptotic limit for $G/G_D$ that depends 
sensitively upon $b$ but is less than unity for reasonable values.
The greatest sensitivity to the shape of the density is found in a
transition region for $Q^2$ that ranges from several tenths to 
several (GeV/$c$)$^2$, depending upon $b$.
Thus, data with similar general features can be fit by modulating
a basic Gaussian with an even polynomial, where the polynomial degree can be
minimized by an optimal choice of $b$.
For $G_{En}$ one need only require the polynomial part of $\tilde{\rho}(k)$
to begin with $k^2$ to ensure that the net charge vanishes.
Expansions of this form are no more complicated than other parametrizations 
in common use, but are free of unphysical cusps at origin.

\section{Analysis procedures}
\label{sec:procedure}

\subsection{Linear expansions}
\label{sec:expansions}

To extract radial densities from the nucleon form factor data we 
employ techniques originally developed for fitting radial distributions 
to data for scattering of electrons or protons from nuclei
\cite{Dreher74,Friar73,Kelly88a}.
Simple models with a small number of parameters do not offer
sufficient flexibility to provide a realistic estimate of the uncertainty
in a radial density.
Rather, we employ linear expansions in complete sets of basis functions 
that are capable of describing any plausible radial distribution without 
strong {\it a priori} constraints upon its shape.
Such expansions permit one to estimate the uncertainties in the fitted
density due to both the statistical quality of the data and the 
inevitable limitation of experimental data to a frequency range,  
$k \leq k_{max}$, where
\begin{equation}
k_{max} = \frac{Q_{max}}{\sqrt{1+\frac{Q_{max}^2}{4m^2}}}
\end{equation} 
is the maximum spatial frequency sampled by experimental data limited
to $Q \leq Q_{max}$.
The uncertainty due to limitation of $k$ is known as 
{\it incompleteness error}.

A radial density can be represented as an expansion of the form
\begin{equation}
\rho(r) = \sum_n a_n f_n(r)
\end{equation}
where the basis functions $f_n(r)$ are drawn from any convenient complete set.
The corresponding Fourier transform then takes the form
\begin{equation}
\tilde{\rho}(k) = \sum_n a_n \tilde{f}_n(k)
\end{equation}
where
\begin{equation}
\tilde{f}_n(k) = \int_0^\infty{ dr \; r^2 j_0(kr) f_n(r) }
\end{equation}
represent basis functions in momentum space.
The expansion coefficients, $a_n$, are fitted to form factor data 
subject to several minimally restrictive constraints to be discussed shortly.
Analyses of this type are often described as model independent because a 
complete basis can reproduce any physically reasonable density; 
if a sufficient number of terms are included in the fitting procedure the
dependence of the fitted density upon the assumptions of the model is 
minimized. 
By contrast, simple parametrizations like the Galster model severely
constrain the shape of the fitted density.

We consider two bases that have been found useful in the analysis of
electron or proton scattering data.
The present discussion is limited to monopole densities, 
but generalizations to higher angular momenta are discussed in 
Refs.\ \cite{Kelly88a,Kelly91b}.  

The Fourier-Bessel expansion (FBE) employs basis functions of the form
\begin{mathletters}
\begin{eqnarray}
f_n(r) &=& j_0(k_n r) \Theta(R_{max}-r) \\
\tilde{f}_n(k) &=& 
\frac{(-)^n R_{max}}{k^2 - k_n^2}j_0(k R_{max})
\end{eqnarray}
\end{mathletters}
where $\Theta$ is the unit step function, $R_{max}$ is the expansion radius,
and $k_n=n\pi/R_{max}$.
One advantage of the FBE is that the contribution of each term to the form
factor is concentrated around its characteristic frequency $k_n$ so that 
a coefficient $a_n$ is largely determined by data with $k \sim k_n$.
The larger the expansion radius $R_{max}$, the smaller the spacing between 
successive $k_n$ and the greater the sensitivity one has to variations in 
the form factor.
One should choose $R_{max}$ to be several times the root-mean-square radius
but not so large that an excessive number of terms is needed to span the 
experimental range of momentum transfer.
Terms with $k_n > k_{max}$ provide an estimate of the incompleteness error.
We chose $R_{max} = 4.0$ fm, but the results are insensitive to its exact 
value.
However, a disadvantage of the FBE is that a relatively large number
of terms is often needed to accurately represent a typical confined density.

Alternatively, the Laguerre-Gaussian expansion (LGE) employs basis functions
of the form
\begin{mathletters}
\begin{eqnarray}
f_n(r) &=& e^{-x^2} L_n^{1/2}(2x^2) \\
\tilde{f}_n(k) &=& \frac{\sqrt{\pi}}{4}b^3 (-)^n e^{-y^2} L_n^{1/2}(2y^2)
\end{eqnarray}
\end{mathletters}
where $x=r/b$, $y=kb/2$, and $L_n^a$ is a generalized Laguerre polynomial.
A significant advantage of the LGE is that the number of terms needed to
provide a reasonable approximation to the density can be minimized by 
choosing $b$ in accordance with the natural radial scale.
We chose $b=0.556$ fm such that the mean-square radius of the Gaussian
factor is consistent with that of the common dipole parametrization
of Sachs form factors.
We then find that the magnitude of $a_n$ decreases rapidly with $n$, 
but the quality of the fit and the shape of the density are actually
independent of $b$ over a wide range.
However, a disadvantage of the LGE is that the basis functions are
not localized in momentum space so that the coefficients tend to
be correlated more strongly than for the FBE.

\subsection{Constraints}
\label{sec:constraints}

The expansion coefficients are obtained by minimizing 
\begin{equation}
\chi^2 = \sum_i \left( \frac{y_i-\bar{y}_i}{\delta y_i} \right)^2
\end{equation}
where $\bar{y}_i$ is the fitted value of a quantity $y_i$ with
uncertainty $\delta y_i$.
In addition to experimental data, 
the set ${y_i}$ generally includes pseudodata used to enforce 
constraints and to estimate the incompleteness error associated with
the limitation of experimental data to a finite range of momentum transfer.

The absence of data for very large $Q^2$ requires some
constraint upon the behavior of $\tilde{\rho}(k)$ for $k_{max} < k < k_{m}$.
Furthermore, inversion of the Fourier transform also requires an assumption
about the experimentally inaccessible region $k>k_m$.
On quite general grounds one expects the asymptotic form factor for a 
confined system to decrease more rapidly than $k^{-4}$
\cite{Friar73};
in particular, this condition ensures that there will be no cusp at 
the orgin.
In fact, our results show that the intrinsic form factor
$\tilde{\rho}(k)$ is well approximated by a Gaussian for large $k$.
Therefore, we will assume that $\tilde{\rho}(k)$ for $k>k_m$ is
bounded by a $k^{-4}$ envelope and use the flexibility afforded by
that envelope to estimate the incompleteness error due to the limitation of
experimental information to the range $k<k_{max}$.
Although some restriction is needed to stabilize the fits, 
the $k^{-4}$ envelope probably overestimates the uncertainties in 
unmeasured form factors and their effect upon uncertainties in  
fitted densities; 
nevertheless, we prefer to employ minimally restrictive
constraints so that those densities will have the best possible 
model independence.

More detailed discussions of the method may be found in  
Refs.\ \cite{Dreher74,Friar73,Kelly88a},  
but the basic idea is to supplement the experimental data by pseudodata
of the form $\tilde{\rho}(k_i) = 0 \pm \delta \tilde{\rho}(k_i)$
whose uncertainties are based upon a reasonable model of the
asymptotic behavior of the form factor for $k_i > k_{max}$
where $k_{max}$ is the spatial frequency corresponding to the 
maximum measured $Q^2$. 
Therefore, uncertainties in the form factor for $k>k_{max}$ are
based upon an envelope of the form   
\begin{mathletters}
\label{eq:envelope}
\begin{eqnarray}
\delta\tilde{\rho}(k) &=& \sqrt{\frac{1}{3}} \; \rho_{lim}(k) \\ 
\rho_{lim}(k) &=& \left| \tilde{\rho}(k_{max}) \right| 
\left( \frac{k_{max}}{k} \right)^4
\end{eqnarray}
\end{mathletters}
where the factor of $\sqrt{1/3}$ represents the variance of a 
uniform distribution of unit width.
When using FBE the pseudodata are chosen at the characteristic
frequencies $k_n = n \pi/R_{max}$ with $n>k_{max}R_{max}/\pi$, while a uniform
spacing of $\Delta k_i = 1.0$ fm$^{-1}$ was employed for LGE.
The error band for a fitted density is computed from the covariance matrix
for the $\chi^2$ fit and includes the incompleteness error.
A detailed discussion of the decomposition of the density uncertainty 
into statistical and incompleteness errors may be found in  
Ref.\ \cite{Kelly88a}.  

Recognizing that pQCD imposes an asymptotic limit of the form  
$G \propto Q^{-4}$ upon the Sachs form factors, one might be tempted
to employ pseudodata for $G$ at large $Q^2$.
If one knew how to estimate the proportionality constant, this 
procedure could be used to regulate $\tilde{\rho}(k)$ for
$k_{max}<k<k_m$ but would not be sufficient for construction of
the radial density because inversion of the Fourier transform
also requires information for the $k>k_m$ region that is
inaccessible to electron scattering.
Although we expect $\tilde{\rho}(k)$ to be small for $k>2m$,
we cannot simply set it to zero because an abrupt cut-off would
introduce unreasonable density oscillations at very large radii. 
The present procedure estimates the uncertainty in the radial
density arising from both the unmeasured and the unmeasurable
ranges of spatial frequency.
In this model, the minimum uncertainty in density is governed by 
the nucleon Compton wavelength and can be interpreted as an
irreducible smearing by zitterbewegung.

Small but undesirable oscillations in fitted densities at large radii
were suppressed using a {\it tail bias} based upon the method
discussed in Ref. \cite{Kelly91b}.
We employed a tail function of the form $t(r) \propto e^{-\Lambda r}$,
based upon the successful dipole parametrization for low $Q^2$, 
and included in the $\chi^2$ fit a penalty function of the form
\begin{equation}
\chi^2_r = \sum_{i=1}^{N_d}{ 
\left( \frac{ \rho(r_i) - t(r_i) }{w \; t(r_i)} \right)^2 }
\end{equation}
to suppress strong deviations from the tail function.
The radial pseudodata were constucted on the grid 
$r_i = r_m + i\Delta r$ for $i=1,N_d$ in the range $r>r_m$.
We chose $\Lambda = 4.27$ fm$^{-1}$, $r_m = 2.0$ fm, $\Delta r = 0.2$ fm,
$N_d = 10$, and $w=2$, 
but the results are rather insensitive to these details. 
The tail bias improves the convergence of moments of the density 
but has little effect upon a fitted density in the region where 
it is large. 

The fitting procedure also permits constraints to be placed on integral
moments of the radial density.  
We define fitted moments by
\begin{equation}
\bar{M}_\alpha = \int_0^\infty dr \; r^{2+\alpha} \rho(r) 
\end{equation}
where here $\rho(r)$ is the fitted density and include a penalty function 
of the form
\begin{equation}
\chi^2_M = 
\left( \frac{M_0 - \bar{M}_0}{\delta M_0} \right)^2 +
\left( \frac{M_2 - \bar{M}_2}{\delta M_2} \right)^2 
\end{equation}
where $M_\alpha$ is the measured value and $\delta M_\alpha$ is its
uncertainty.
The constraint on the neutron charge was enforced by means of a
pseudodatum $M_0 = 0 \pm 10^{-6}$.
In addition, the atomic physics datum for $M_2$ from Ref.\ \cite{Kopecky97} 
was included in fits made to the neutron charge density. 

It is also useful to define a fitted transition radius $\bar{R}$ as
\begin{equation}
\label{eq:Rtr}
\bar{R} = \sqrt{ \bar{M}_2 / \bar{M}_0 }
\end{equation} 
for $G_{Ep}$, $G_{Mp}$, or $G_{Mn}$.
Thus, the fitted transition radius is correlated with the 
experimental normalization at low $Q^2$.
If the fitted $\bar{M}_0$ were constrained, the uncertainty in $\bar{R}$ 
would be artificially reduced.
Therefore, no constraints were placed on $M_0$ for $G_{Ep}$, $G_{Mp}$, or 
$G_{Mn}$ ---
given that those intrinsic densities were defined with unit normalization,
the fitted values of $\bar{M}_0$ test the normalization of the experimental
data.

\subsection{Data selection}
\label{sec:data}

We tried to select the best available data in each range of $Q^2$, 
with an emphasis upon recent data using recoil or target polarization 
wherever available.
These selections are summarized in Table \ref{table:data}.
Although a thorough review of the data for nucleon electromagnetic form 
factors is beyond the scope of the present work, in this section we
provide brief explanations for some of our selections and omissions.

$G_{Mp}$ data were taken from the compilation of 
H\"ohler {\it et al}.\ \cite{Hohler76}
for $Q^2 < 0.15$ (GeV/$c$)$^2$ and for larger $Q^2$ from the compilation
and re-analysis made by Brash {\it et al}.\ \cite{Brash02} to improve
the correction for the small contribution of $G_{Ep}$ to cross section 
at large $Q^2$.
Values for $G_{Ep}$ were obtained by multiplying the recent recoil 
polarization measurements of $G_{Ep}/G_{Mp}$ from
Refs.\ \cite{MKJones00,Gayou02,Pospischil01,Milbrath98}
by the Brash parametrization of $G_{Mp}$.
Supplementary recoil polarization data from Ref.\ \cite{Gayou01} 
were omitted --- those data are consistent with those selected but
have larger statistical uncertainties.
In addition, the cross section data for $G_{Ep}$ from  
Refs.\ \cite{Simon80,Price71} were used at low $Q^2$.
The cross section data for $G_{Ep}$ at $Q^2 > 1$ (GeV/$c$)$^2$
from Refs.\ \cite{Walker94,Andivahis94} were omitted because, 
as shown in Ref.\ \cite{MKJones00}, they are in significant disagreement
both with the recoil polarization data and with each other
presumably because the Rosenbluth technique becomes increasingly difficult 
as $Q^2$ increases and the relative contribution of the electric form factor
to the unpolarized cross section becomes quite small.
Nevertheless, we eagerly await new results from a proposed improvement
of the Rosenbluth method \cite{JLAB01-001}. 

The neutron magnetic contribution is large enough at high $Q^2$ to
employ quasielastic electron-deuteron scattering with subtraction 
of the proton contribution.
Data of this type were obtained from Refs.\ \cite{Lung93,Rock82}.
At low $Q^2$ the model dependence of the quasielastic method becomes
relatively large.
Markowitz {\it et al.}\ \cite{Markowitz93} measured the quasifree
neutron knockout cross section for the $d(e,e^\prime n)$ reaction and
calibrated the efficiency of the neutron detector using associated
particle production in the deuteron photodisintegration reaction, 
$d(\gamma,pn)$.
The dependence upon the deuteron wave function can be reduced
by analyzing the ratio between quasifree cross sections for neutron
or proton knockout, indicated as the $d(e,e^\prime N)$ reaction
in Table \ref{table:data}, with relatively small corrections made
for meson-exchange currents and final-state interactions (FSI).
Bruins {\it et al.}\ \cite{Bruins95} calibrated their neutron
detector using the $p(\gamma,\pi^+ n)$ reaction while 
Refs. \cite{Anklin94,Anklin98,Kubon02} employed elastic 
neutron-proton scattering.
The associated-particle techniques permit calibration {\it in situ}
but must correct the bremsstrahlung measurements for the contributions
of three-body reactions from electroproduction that lie outside the
acceptance \cite{Jourdan97,Bruins97},
whereas the $p(n,p)n$ reaction is kinematically complete but requires 
calibration at a different facility and under different conditions than
used for the reaction of interest.
Unfortunately, these methods remain in substantial disagreement;
because we are not convinced there is a compelling preference, we
include the data from both methods in the present analysis.
Finally, we also include recent low $Q^2$ data \cite{Xu00} from inclusive 
electron scattering from transversely polarized $^3\vec{{\rm He}}$ that uses 
the Fadeev calculations of Refs.\ \cite{Golak95,Golak01} to correct for 
nuclear structure effects.
These data are more consistent with the coincidence ratio method calibrated
by elastic scattering than by associated-particle production.

We do not use any elastic or quasielastic cross section data for $G_{En}$ 
because the uncertainties arising from nuclear structure are prohibitively 
large.
Polarization techniques offer a signal that is linear in $G_{En}$
and with less model dependence.
Nevetheless, at low $Q^2$ it remains important to correct recoil
polarization data for final-state interactions and target polarization
for nuclear structure.
Most of the data for deuterium targets have been analyzed using the
calculations of Arenh\"ovel {\it et al.} \cite{Arenhovel88,Arenhovel92}
to correct for nuclear structure; in Table \ref{table:data} we cite both
the experimental paper and the subsequent analysis.
The result from the first experiment of this type \cite{Eden94b} has not
been corrected, but the statistical uncertainty was large.
The $Q^2 = 0.4$ (GeV/$c$)$^2$ data for $^3\vec{{\rm He}}(\vec{e},e^\prime n)$
have been analyzed using Fadeev calculations \cite{Golak01}
but Rohe {\it et al.}\ \cite{Rohe99} argue that the plane-wave impulse
approximation (PWIA) is adequate at larger $Q^2$.
We also include values for $G_{En}$ extracted by
Schiavilla and Sick \cite{Schiavilla01}
from an analysis of the deutron quadrupole form factor
obtained from tensor polarization measurements of elastic scattering
\cite{Garcon94,Ferro-Luzzi96,Bowhuis99,Abbott00}.
Although the model uncertainties remain somewhat large, 
this analysis covers a larger range of $Q^2$ and appears to be more
accurate than the older analysis of deuteron elastic scattering
by Platchkov {\it et al.} \cite{Platchkov90}, which is omitted.

Finally, the neutron mean-square charge radius is related to the
neutron-electron scattering length, $b_{ne}$, by
\begin{equation}
\langle r^2 \rangle_n = \frac{3 \hbar}{\alpha m_n c} b_{ne}
\end{equation}
Unfortunately, the measurements are rather difficult and most
techniques require substantial corrections for effects for which
there is often insufficient information.
A recent review of these measurements has been made by 
Alexandrov \cite{Aleksandrov99}, who finds that
most modern measurements cluster around two values.
From measurements of the energy dependence for the transmission of thermal 
neutrons through liquid $^{208}$Pb,
Kopecky {\it et al.} Ref. \cite{Kopecky97} obtained
$b_{ne} = (-1.33 \pm 0.027 \pm 0.03) \times 10^{-3}$ fm, corresponding
to $\langle r^2 \rangle_n = -0.115 \pm 0.003$ fm$^2$. 
This result agrees well with similar measurements by 
Koester {\it et al.}\ \cite{Koester95} for lead isotopes and $^{209}$Bi
and with the results of Krohn and Ringo \cite{Krohn73} using 
the angular distribution for neutron scattering by noble gases.
Alternatively, Alexandrov {\it et al.}\ \cite{Alexandrov85} obtained
$b_{ne} = (-1.60 \pm 0.05) \times 10^{-3}$ fm, corresponding to
$\langle r^2 \rangle_n = -0.138 \pm 0.004$ fm$^2$, 
using neutron diffraction from single crystals of $^{186}$W.
This result is consistent with a bismuth transmission experiment that was also
performed at Dubna, but disagrees by about 5 standard deviations from
the Garching, Argonne, and Oak Ridge experiments.
This discrepancy has been attributed to resonance corrections 
\cite{Leeb93,Alexandrov94} but remains controversial.
The extracted scattering length is strongly correlated with the resonance
correction.
Leeb and Teichtmeister \cite{Leeb93} argue that the correction employed
by Alexandrov {\it et al.}\ \cite{Alexandrov85} requires implausibly large 
contributions from negative energy levels.
Alexandrov \cite{Alexandrov94} argues that the energy-independent resonance
correction should be fitted to the data and that in the absence of definitive
information a negative contribution cannot be excluded. 
Futhermore, neutron diffraction from single crystals of $^{186}$W provides 
a larger signal than the energy dependence of the total cross section, 
and is hence is less sensitive to this correction.
Although this controversy has not yet been resolved satisfactorily,
we decided to employ the most recent result from Oak Ridge,
namely $\langle r^2 \rangle_n = -0.115 \pm 0.003$ fm$^2$,
as a datum in our fit of the neutron charge density and to omit
the Dubna result.
The sensitivity to this choice is discussed in Sec.\ \ref{sec:neutron-radius}.

\section{Results}
\label{sec:results}

\subsection{Form Factors}
\label{sec:results-ff}

\begin{figure}[htbp]
\centerline{ \strut\psfig{file=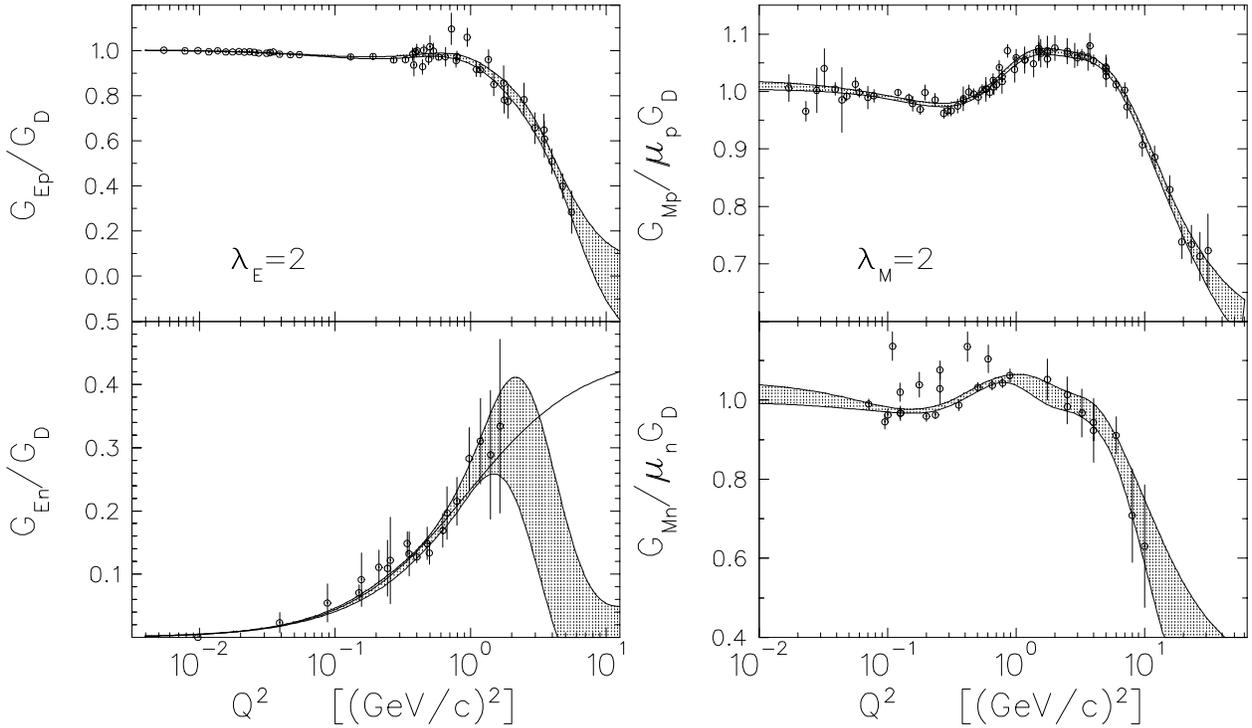,width=6.5in,angle=90} }
\caption{The bands show fits to selected data for nucleon electromagnetic 
form factors using the LGE parametrization with $\lambda_E=\lambda_M=2$.
For $G_{En}$ the solid line shows a two-parameter fit based upon the
Galster parametrization.}
\label{fig:ff}
\end{figure}

Fits to the form factor data are shown in Fig.\ \ref{fig:ff} as bands 
that represent the uncertainties in the fitted form factors.
These bands were computed using the covariance matrix.  
The fits shown in Fig.\ \ref{fig:ff} employ the LGE parametrization with 
$\lambda_E=\lambda_M=2$, but the results using the FBE parametrization
are practically indistinguishable.
Nor do these fits depend upon the choices for $b$, $R_{max}$, or
details of the constraints.
Fits using $\lambda<2$ are almost identical within the ranges spanned by
experimental data, but their error bands grow more rapidly at larger $Q^2$.
The rapidly decreasing dipole form factor is divided out to emphasize 
the deviations at large $Q^2$ from this characteristic behavior. 
For $G_{En}$ we also display a simple two-parameter fit using the 
Galster parametrization.

\begin{figure}[hbtp]
\centerline{ \strut\psfig{file=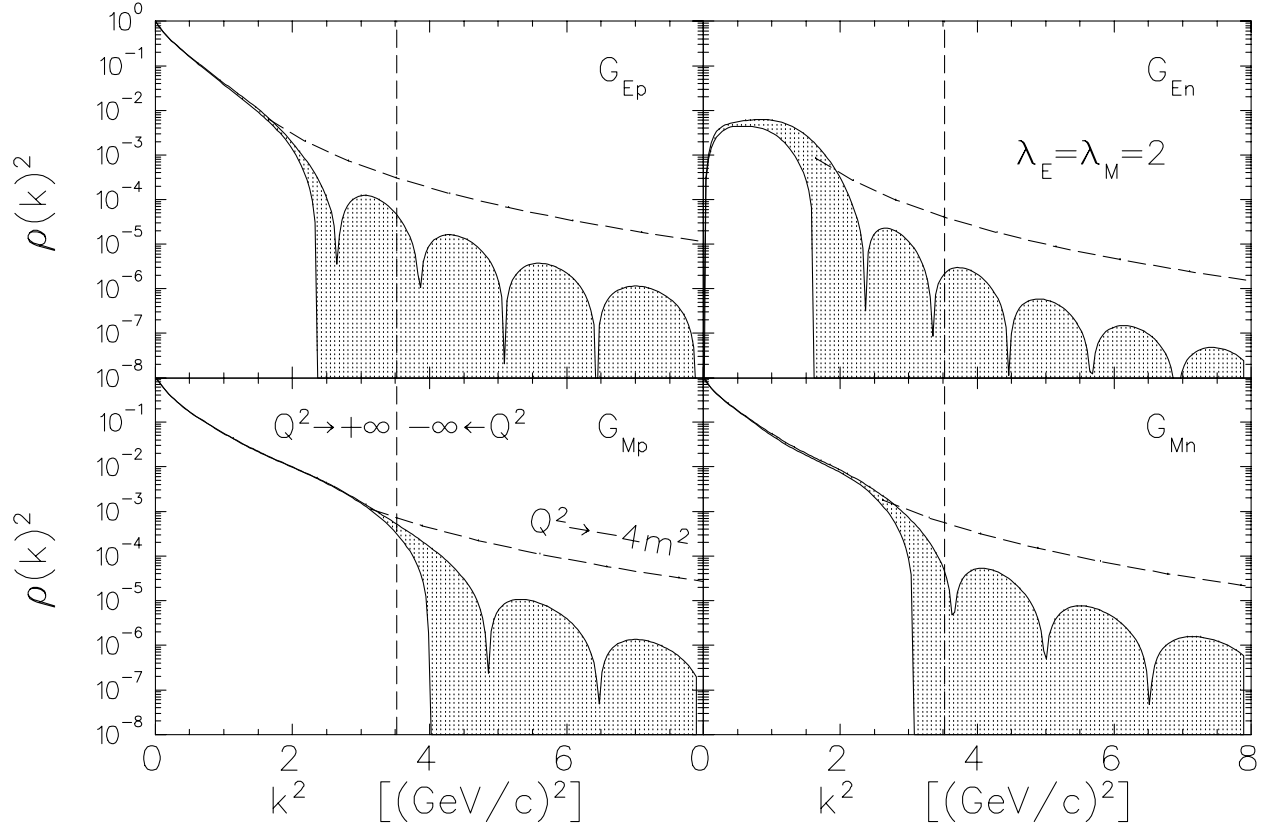,width=6.5in,angle=90} }
\caption{Fourier transforms of the nucleon charge and magnetization
densities are shown as error bands.  
The LGE parametrization was used with $\lambda_E=\lambda_M=2$.
The dashed lines show the upper limits used for estimation of 
incompleteness errors.
The vertical lines divide the regimes of spacelike and timelike
$Q^2$, where the timelike threshold, $Q^2 < -4m^2$, is approached
in the limit $k^2 \rightarrow \infty$.
Note that electron scattering is limited to the spacelike regime,
wherein $Q^2 \rightarrow \infty$ is represented by $k \rightarrow 2m$.}
\label{fig:rhok}
\end{figure}

The intrinsic form factors, obtained via the relativistic
transformation prescribed by Eq. (\ref{eq:rhok}),
are shown in Fig.\ \ref{fig:rhok} using $\lambda_E=\lambda_M=2$.
From these figures we observe that for moderate $k^2$
three of the four intrinsic form factors resemble simple Gaussians, 
while the intrinsic neutron charge form factor requires an 
additional factor of $k^2$ in first approximation.
Consequently, only a few terms of the Laguerre-Gaussian expansion 
are needed to obtain good fits, with higher-order terms used 
primarily for the estimation of the incompleteness error.
Although it is possible to obtain fairly good fits using just 2 
terms for $G_{En}$, 4 for $G_{Ep}$, or 6 for $G_{Mp}$ and $G_{Mn}$,
in order to minimize model dependence and to evaluate incompleteness
errors we employed 20 terms for each of the four form factors.

The widths of the form factor bands are governed by the quality 
of the experimental data in the ranges of $k^2$ where data are 
available and for larger $k^2$ by the asymptotic envelopes 
indicated by dashed curves.
Note that the uncertainties in the fitted form factors for $k>k_{max}$ 
are reduced by the factor of $1/3$ used in Eq.\ (\ref{eq:envelope}) 
to transform from a uniform to a normal distribution and by the effect 
of constraints upon densities at large radii.
Although the intrinsic form factors fitted to data for $G_{Ep}$, 
$G_{Mp}$, and $G_{Mn}$ appear to decrease more rapidly than the 
$k^{-4}$ envelopes, 
we prefer to employ these more generous uncertainties rather than 
to impose the steeper declines suggested by extrapolation from the
measured into the unmeasurable region (where $k>k_m$).
The use of steeper envelopes would simply reduce the uncertainties 
in the extracted densities without affecting their central values.
Therefore, $k^{-4}$ envelopes were matched to fitted form factors 
at  $k_{max} = 6.5$, 9.0, and 8.2 fm$^{-1}$ for $G_{Ep}$, $G_{Mp}$,
and $G_{Mn}$ based upon the experimental $Q_{max}$ for each
form factor.
However, uncritical application of the same procedure to $G_{En}$
would suppress the high-frequency components of its intrinsic
form factor too strongly because $Q_{max}$ for $G_{En}$ is  
presently too small to expect $\tilde{\rho}(k)$ to decrease more
rapidly than $k^{-4}$.
Fig.\ \ref{fig:ff} shows that the data presently available for
$G_{En}$ are compatible with the Galster parametrization,
but the procedure used for the other form factors would cause
$G_{En}/G_D$ to decrease fairly rapidly beyond the range of
these data.
On the other hand, it is reasonable to expect $G_{En}/G_D$ to 
decrease for $Q^2$ beyond a few (GeV/$c$)$^2$, 
as observed in the other form factors.
Therefore, in order to permit the positive slope for 
$G_{En}/G_D$ to continue over a limited but larger range of $Q^2$,
we used the same value of $k_{max}$ for both $G_{En}$
and $G_{Ep}$ even though the $G_{En}$ data are limited to
$k < 5.4$ fm$^{-1}$.
We believe that this compromise provides a more reliable
extrapolation to higher $Q^2$ and that the increased estimate
of incompleteness error is more realistic, but obviously it is
very important to acquire accurate data for $G_{En}$ at
higher $Q^2$.

\begin{figure}[ht]
\centerline{ \strut\psfig{file=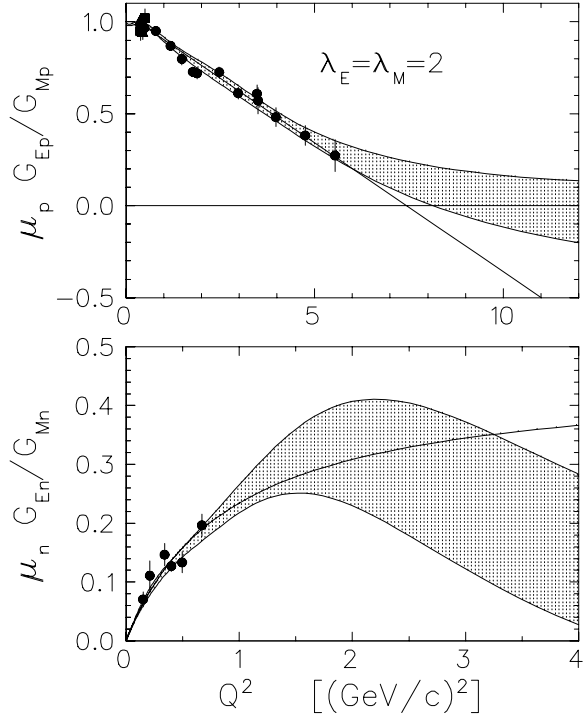,width=3.0in,angle=90} }
\caption{Comparison between data for $G_E/G_M$ obtained from
polarization measurements with fits made to the entire data sets
employed for nucleon electromagnetic form factors.
Results for the LGE parametrization $\lambda_E=\lambda_M=2$ are shown as
bands.
Also shown are the linear parametrization proposed by 
{\protect\cite{MKJones00}} for the proton and a fit based upon
the Galster parametrization for the neutron. }
\label{fig:gratio_lge2}
\end{figure}

These fits to intrinsic magnetic form factors do not change sign 
within the experimentally accessible region, $k<k_m$, but the
fitted proton intrinsic charge form factor suggests a node between
the present experimental limit, $k_{max} = 6.5$ fm$^{-1}$, and
ultimate limit, $k_m=9.5$ fm$^{-1}$.
Consequently, this model suggests a zero crossing in $G_{Ep}$
near $Q^2 \sim 10$ (GeV/$c$)$^2$.
Figure \ref{fig:gratio_lge2} compares the form factor ratio
$\mu G_{E}/G_{M}$ deduced from the fitted form factors with
experimental data using recoil polarization for the proton
or using either recoil or target polarization for the neutron.
For the proton we also show the linear parametrization proposed by 
Jones {\it et al.}\ \cite{MKJones00} for $Q^2 > 0.3$ (GeV/$c$)$^2$,
while for the neutron we show a new fit using the Galster parametrization,
Eq.\ (\ref{eq:Galster}), that gave $A=0.90 \pm 0.02$ and $B=3.8 \pm 0.5$.
The data for the proton do not distinguish between linear and LGE 
parametrizations, but according to pQCD one would expect 
$G_{Ep}/G_{Mp}$ to approach a constant for sufficiently large $Q^2$.
Extrapolation of the LGE parametrization suggests that the
asymptotic ratio will be very small, but data at much larger $Q^2$
are needed to establish that level.
An extension to 9 (GeV/$c$)$^2$ has been approved \cite{JLAB01-109},
but larger $Q^2$ remains desirable.
Similarly, the present data for $G_{En}/G_{Mn}$ are compatible 
with the Galster parametrization but remain limited to rather small $Q^2$.
Consequently, the extrapolation to larger $Q^2$ is rather uncertain.
If an approved experiment using the $^3\vec{\rm He}(\vec{e},e^\prime n)$
reaction \cite{JLAB02-013} achieves the proposed $\pm13\%$ statistical 
uncertainty at $Q^2 = 3.4$ (GeV/$c$)$^2$, the error band will be reduced 
to about the same width and the extrapolation much improved.
Nevertheless, there is little reason to expect the asymptotic limit
to be reached earlier for the neutron than for the proton.

Although a review of recent theoretical calculations is beyond the scope
of the present work, it is probably worth mentioning a few which
describe the new $G_{Ep}/G_{Mp}$ data relatively well.
Among these the earliest is the chiral soliton model of 
Holzwarth \cite{Holzwarth96}, which predicted the linear descrease
with respect to $Q^2$ and a sign change near 10 (GeV/$c$)$^2$.
More recently \cite{Holzwarth02}, modifications of the vector meson 
parameters were made to improve the fits to the neutron form factors, 
but the ratio $G_{Mn}/G_{Mp}$ is not reproduced.
Furthermore, because the chiral soliton model uses $\lambda_E=0$ and
$\lambda_M=1$, Holzwarth found it necessary to artificially increase
the soliton mass in order to obtain reasonable fits at large $Q^2$.
Alternatively, Lu {\it et al.}\ \cite{Lu00,Lu98} obtained a good fit
to the $G_{Ep}/G_{Mp}$ data for $Q^2 \lesssim 3$ (GeV/$c$)$^2$ 
by adjusting the bag radius in the cloudy bag model, but the ratio
appears to level off well above the more recent data for higher $Q^2$.
Note that this model uses $\lambda_E=\lambda_M=1$.
The covariant calculation of Boffi {\it et al.}\ \cite{Boffi02} using
the point-form spectator approximation provides reasonably accurate
predictions of the form factors for $Q^2 \lesssim 5$ (GeV/$c$)$^2$,
although there remains a significant discrepancy for $G_{Mp}$ near
the end of this range.
The light-front calculations of Cardarelli and Simula \cite{Cardarelli00} 
using one-gluon exchange
and the light-cone diquark model of Ma {\it et al.}\ \cite{Ma02} 
also reproduce the linear $Q^2$ dependence of $G_{Ep}/G_{Mp}$ fairly
well.

\subsection{Densities}
\label{sec:results-densities}

\begin{figure}[htbp]
\centerline{ \strut\psfig{file=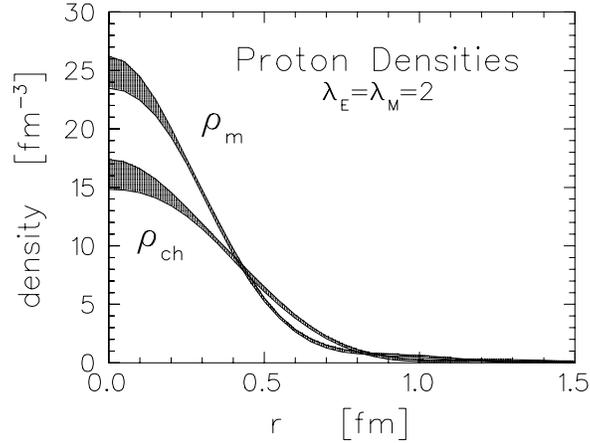,width=3.0in,angle=90} }
\caption{Comparison between charge ($\rho_{ch}$) and 
magnetization ($\rho_m$) densities for the proton fitted using
the LGE parametrization with $\lambda_E=\lambda_M=2$.  
Both densities are normalized to $\int dr \; r^2 \rho(r) = 1$.}
\label{fig:proton}
\end{figure}

Proton charge and magnetization densities are compared in 
Fig.\ \ref{fig:proton}.
Both densities are measured very precisely, 
with uncertainties at the origin better than 6\% for magnetization 
or 8\% for charge.
Incompleteness dominates in the interior region while statistical
errors become comparable in the surface region.
As shown by the variation of $G_{Ep}/G_{Mp}$ in the top panel of 
Fig.\ \ref{fig:gratio_lge2},
the new recoil-polarization data for $G_{Ep}$ decrease more rapidly
than either the dipole form factor or the magnetic form factor
for $Q^2 > 1$ (GeV/$c$)$^2$.
Consequently, we find that the charge density is significantly softer 
than the magnetization density of the proton.
The densities obtained using LGE or FBE parametrizations are practically
indistinguishable and are independent of the choice of $b$ or $R_{max}$
over wide ranges.
These densities are similar to the Gaussian densities one might expect
from a quark model and are more realistic than the exponential density that 
results from naive nonrelativistic inversion of the dipole form factor.

\begin{figure}[hbtp]
\centerline{ \strut\psfig{file=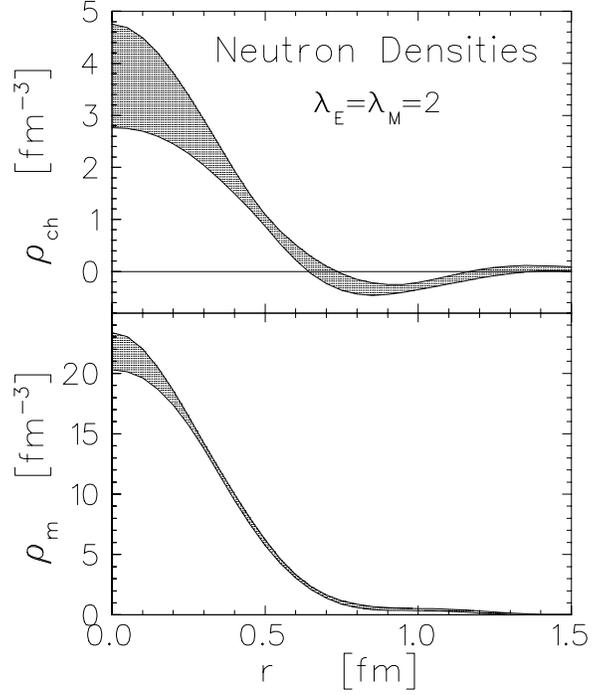,width=3.0in} }
\caption{Charge ($\rho_{ch}$) and magnetization ($\rho_m$) densities 
for the neutron 
fitted using the LGE parametrization with $\lambda_E=\lambda_M=2$.}
\label{fig:neutron}
\end{figure}

Neutron densities are shown in Fig. \ref{fig:neutron}.
We find that the magnetization density for the neutron is very similar to 
that for the proton,
although the interior precision is not as good because the range of $Q^2$ 
is smaller and the experimental uncertainties larger.
Limitations in the range and quality of the $G_{En}$ data presently
available result in a substantially wider error band for the neutron
charge density.
Data at higher $Q^2$ are needed to improve the interior precision,
but a useful measurement of the interior charge density is obtained 
nonetheless.
The positive interior density is balanced by a negative surface lobe.
Note that polarization measurements are sensitive to the sign of the density.

\begin{figure}[hbtp]
\centerline{ \strut\psfig{file=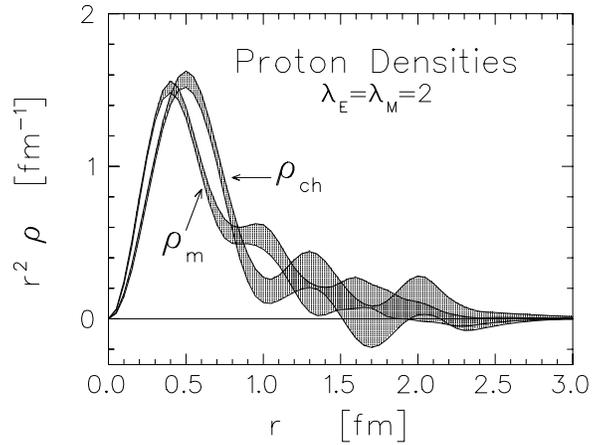,width=3.0in,angle=90} }
\caption{Comparison between proton charge and magnetization densities
using a factor of $r^2$ to emphasize the surface and tail regions.
The fits used the LGE parametrization with $\lambda_E=\lambda_M=2$.}
\label{fig:proton_density2_lge2}
\end{figure}

\begin{figure}[hbtp]
\centerline{ \strut\psfig{file=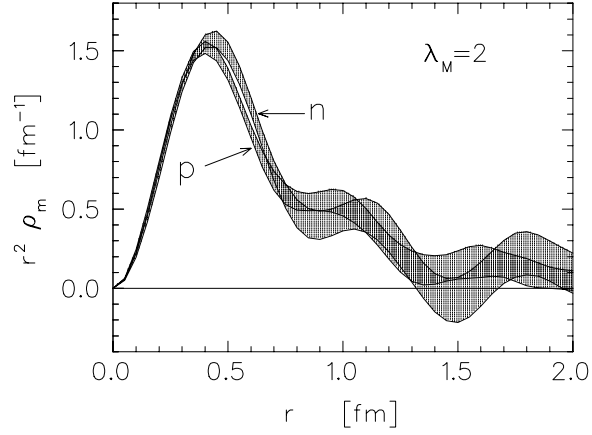,width=3.0in,angle=90} }
\caption{Comparison between neutron and proton magnetization densities
using a factor of $r^2$ to emphasize the surface region.  
The fits used the LGE parametrization with $\lambda_M=2$.}
\label{fig:magnetization}
\end{figure}

Whereas Figs.\ \ref{fig:proton}-\ref{fig:neutron} emphasize the interior 
densities,
it is also of interest to compare these densities in the surface and tail 
regions.
Figures \ref{fig:proton_density2_lge2}-\ref{fig:magnetization} use a factor 
of $r^2$ to emphasize these surface and tail densities. 
Although the densities are small, the reduced
slopes seen between 1 and 1.5 fm in the neutron magnetization and in both 
the charge and the magnetization densities for the proton are seen as 
significant peaks in $r^2 \rho$. 
Virtually identical features also emerge using the FBE parametrization.
These features are independent of $b$ for the LGE or $R_{max}$ for the FBE
parametrization over wide ranges. 
Many attempts were made to suppress structures in $r^2 \rho$ in the
1-1.5 fm region by limiting the number terms in the expansions or by
application of stricter tail biases, but all modifications which did
produce smoother $r^2 \rho$ curves in this region damaged the fits to 
the form factor data for  $Q^2 \gtrsim 1$ (GeV/$c$)$^2$.
Although it is difficult to prove that smoother fits do not exist, 
especially if one is willing to tolerate a moderate increase in $\chi^2$,
we were unable to produce acceptable fits without some structure in
$r^2 \rho$ in this region.
On other hand, because local errors in momentum space can introduce 
artificial oscillations at large radii, 
we did apply an exponential tail bias for $r>2$ fm where little structure 
is expected.
Thus, the smaller oscillations for $r>1.5$ fm are generally consistent
with zero and can be suppressed using the tail bias with little effect 
upon the fits.
We believe that the 2 fm matching radius is sufficiently large to have
minimal influence upon densities for intermediate distances governed by 
data with $Q^2$ of order several (GeV/$c$)$^2$.

The relatively small differences between $G_{Mn}$ and $G_{Mp}$ seen in
Fig.\ \ref{fig:ff} produce the small differences between neutron and
proton magnetization densities shown in Fig.\ \ref{fig:magnetization}.
The peak of $r^2 \rho_m$ is found at a slightly larger radius for
the neutron than for the proton because the form factor decreases a
little more rapidly with respect to $Q^2$.
The secondary peaks in the 1-1.5 fm region are also similar.
Again, these comparisons are independent of the details of the analysis
and are virtually identical using either LGE or FBE parametrizations.

Therefore, the secondary $r^2 \rho$ peaks in the 1-1.5 fm region appear 
to be essential features of the data rather than artifacts of analyses 
based upon linear expansions.
While it is not possible to determine the physical mechanism for such
features from data analysis alone, there is at least one simple candidate.
The tensor interaction between quarks is expected to produce a small $D$-state
component peaking at larger radius than the dominant $S$-state configuration,
and the superposition of these components could yield a secondary peak
at relatively large radius.

\subsection{Fitted Moments}
\label{sec:fitted_moments}

Moments and $\chi^2$ for each fit are listed in Table \ref{table:moments}.
The expansion coefficients require too much space to list but are 
available by request; also note that accurate reproduction of the
error bands would require full covariance matrices.
Here we quote $\chi^2/N$, where $N$ is the number of data points, 
because counting the number of degrees of freedom is not so clear 
when both high-$k$ and large-$r$ constraints are applied.
For each form factor we find that the 6 fits obtained for 3 possible
choices of $\lambda$ used in either LGE or FBE parametrizations are
essentially identical and give the same values of $\chi^2/N$.
The normalization for $G_{Ep}$ is consistent with unity 
within the 0.5\% systematic uncertainty claimed by
Simon {\it et al.} \cite{Simon80} for their data at low $Q^2$.
For the $G_{Mp}$ at low $Q^2$ we employed the results of
H\"ohler {\it et al.} \cite{Hohler76} who adjusted the relative
normalizations of several data sets to a common standard.
The normalization produced by the present fit is consistent 
with the systematic uncertainty in that standard.
Except for $G_{Mn}$, data selected from several sources appear to be
mutually consistent and the quality of the fitted form factors is
very good.
Although the low-$Q^2$ data for $G_{Mn}$ have improved in recent years,
significant systematic discrepancies remain.
Recent data from Refs.\ \cite{Xu00,Anklin94,Anklin98,Kubon02}
with small statistical uncertainties suggest a small dip near 0.2 
and a peak near 1 (GeV/$c$)$^2$.
However, the data from Refs.\ \cite{Markowitz93,Bruins95} are 
inconsistent with the fit and inflate $\chi^2$.
Nor do the data for $G_{Mn}$ reach sufficiently low
$Q^2$ to strongly constrain the normalization and the data sets
are not entirely consistent either;
consequently, the present analysis suggests that a 2\% 
normalization error remains.
We decided to retain the data from  
Refs.\ \cite{Markowitz93,Bruins95},
despite their deviation from the fit because we are not
entirely convinced of the transportability of the efficiency
calibration from a hadron facility to an electron facility.

The final two columns of Table \ref{table:moments} 
list the Sachs radius, $\xi$, defined by Eq. (\ref{eq:xi}) 
in terms of the  initial logarithmic derivative of the Sachs form
factor and the transition radius, $\bar{R}$, obtained from integral moments 
of the density fitted for a specific $\lambda$ according to Eq. (\ref{eq:Rtr}).
The dependencies of rms radii for $G_{Ep}$, $G_{Mp}$, and
$G_{Mn}$ upon $\lambda$ are consistent with Eq. (\ref{eq:rms}),
showing a significant discrete ambiguity arising from the
model dependence of the form factor to density inversion.
By contrast, for each of these form factors all 6 determinations 
of $\xi$ are consistent with each other, demonstrating
that the fitted $\xi$ is a model-independent property of the 
Sachs form factor.

\subsubsection{Proton charge radius}
\label{sec:proton_radius}

The analysis of Simon {\it et al.} \cite{Simon80} has been accepted 
for about two decades as the definitive determination of the proton charge 
radius, but more recently there has been renewed interest in that quantity 
now that the finite-size corrections have become the dominant uncertainty
in theoretical calculations of the 1S Lamb shift in hydrogen.
For example, Melnikov and van Ritbergen \cite{Melnikov00} argue that
the Lamb shift provides the most accurate measurement of the proton
charge radius and deduced a value 0.883(14) fm that is somewhat larger 
than the 0.862(12) fm obtained by Simon {\it et al}.
Although our definition for the intrinsic charge radius depends
upon the choice of $\lambda_E$ employed to fit the Sachs form factor, 
the definition generally employed by QED theorists corresponds to
the quantity we labelled as $\xi_{Ep}$ that is based upon the initial 
slope of $G_{Ep}(Q^2)$ and is independent of $\lambda_E$.
Thus, it is the fitted value of $\xi_{Ep}$ that should be compared
with the Lamb shift result.

\begin{figure}[hbtp]
\centerline{ \strut\psfig{file=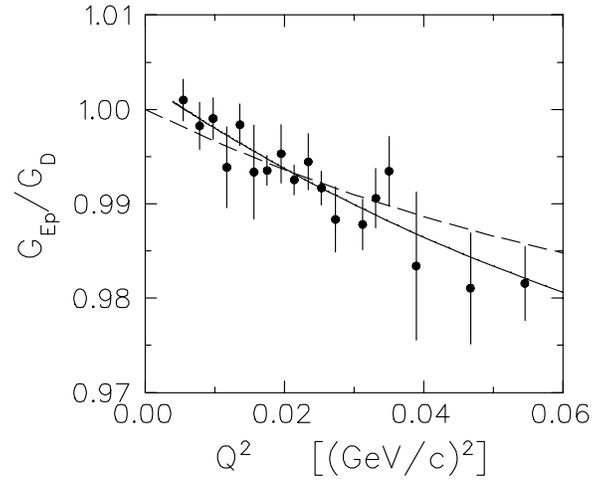,width=3.0in,angle=90} }
\caption{Low $Q^2$ fits to proton charge radius.  
The solid line shows our LGE fit with $\lambda_E=0$ to the entire data
set while the dashed line shows the monopole parametrization of
Simon {\it et al.}\ {\protect\cite{Simon80}}.}
\label{fig:rp}
\end{figure}

Our fit using the LGE parametrization is compared with the data from
Simon {\it et al.} in Fig.\ \ref{fig:rp}.
Also shown is the monopole fit made by Simon {\it et al}.\
to the data for $Q^2 < 2.3$ (GeV/$c$)$^2$ that were available at that time.
That analysis gave a smaller value, $\xi_{Ep} = 0.862 \pm 0.012$ fm,
but does not fit the low $Q^2$ data as well as our LGE fit. 
Our fit employs the entire data set described in Sec.\ \ref{sec:data}
even though  Fig.\ \ref{fig:rp} shows only the lowest $Q^2$ region.
Fits made using different values of $\lambda_E$ or using the FBE
parametrization are indistinguishable and give values for $\xi_{Ep}$ 
that are consistent within their quoted uncertainties.
Therefore, we claim that $\xi_{Ep} = 0.88 \pm 0.01$ fm
represents a model-independent property of the experimental data
even if its interpretation as a charge radius depends upon the choice 
of $\lambda_E$. 
This value is consistent with the rms radius derived by 
Melnikov and van Ritbergen \cite{Melnikov00} from the Lamb shift.

Coulomb distortion may also affect the charge radius obtained by
electron scattering.
Rosenfelder \cite{Rosenfelder00} analyzed this effect using a
distorted wave Born calculation to correct measured electron
scattering cross sections for Coulomb distortion, thereby obtaining
effectively plane-wave cross sections.
The charge radius was then obtained by fitting the resulting
form factors with low-order polynomials.
Although the form factor corrections were typically less than 1\%, 
the value of $\xi_{Ep}$ extracted from the adjusted data of 
Simon {\it et al}.\ nevertheless increases by about 0.008 -- 0.013 fm 
depending upon the fitting strategy and the degree of the polynomial.
However, that analysis employed a Coulomb potential obtained from a
charge density of the form
\begin{equation}
\rho(r) = \int \frac{d^3 k}{(2\pi)^3} e^{i {\bf k} \cdot {\bf r}}
\frac{G_E(k^2)}{\sqrt{1+\frac{k^2}{4m^2}}} 
\end{equation}
This form not only lacks the Lorentz-contraction factor for spatial 
frequency, but uses a value for $\lambda_E = -1/2$ that is inconsistent
with the high $Q^2$ behavior of the form factor.
The question of Coulomb distortion of the low $Q^2$ form factors 
merits further investigation, 
but a consistent relativistic relationship between form factor 
and density must be employed.
Nevertheless, the magnitude of that correction appears to be smaller than
the uncertainty in the fitted quantity.

\subsubsection{Proton magnetization radius}
\label{sec:proton_mag_radius}

With $\xi_{Mp} = 0.85 \pm 0.03$ fm the proton magnetization
radius appears to be slightly smaller than its charge radius,
as expected from Fig.\ \ref{fig:proton},
but the uncertainty is as large as the difference because
the data at very low $Q^2$ are less precise for $G_{Mp}$ than
for $G_{Ep}$.
A substantial part of the uncertainty in $\xi_{Mp}$ is due
to the uncertainty in normalization.  
If we constrain $M_0$ to unity, the changes in
fitted form factors and densities are relatively small, 
but we obtain values for 
$\xi_{Ep} = 0.862 \pm 0.006$ fm and
$\xi_{Mp} = 0.835 \pm 0.006$ fm
that appear to be much more precise.
The constrained fit to $G_{Ep}$ is close to the result
of Simon {\it et al.} shown in Fig.\ \ref{fig:rp}
at very low $Q^2$.
This analysis demonstrates that there is an appreciable
difference between the proton charge and magnetization
densities, but it also highlights the importance of
precise absolute normalization at very low $Q^2$.

\subsubsection{Neutron magnetization radius}
\label{sec:neutron_radius}

Our fits to the $G_{Mn}$ data give a value for 
$\xi_{Mn} = 0.92 \pm 0.07$ fm 
that has substantial uncertainty because the lack of data for very low 
$Q^2$ permits significant widening of the error band as 
$Q^2 \rightarrow 0$.
By contrast, Kubon {\it et al.}\ \cite{Kubon02} obtained a value 
$\xi_{Mn} = 0.873 \pm 0.011$ fm 
that appears to be much more precise.
However, their continued-fraction parametrization automatically
constrains the normalization at $Q^2=0$.
If we constrain the normalization by requiring $M_0 = 1$, then
the LGE analysis gives a value for $\xi_{Mn} = 0.881 \pm 0.018$ fm
that agrees with Kubon {\it et al}.
Note that we included data from Refs.\ \cite{Markowitz93,Bruins95}
that were omitted by Kubon {\it et al}.\ and that deviate strongly 
from our LGE fit, but these data appear to have little influence 
upon the fitted normalization and radius.
Although the fitted normalization is consistent with unity,
the 3\% uncertainty does have an appreciable effect upon the
uncertainty in $\xi_{Mn}$.
We chose not to constrain the normalization in the standard
analysis because the systematic errors in neutron efficiency 
have been a big problem for $G_{Mn}$ measurements and there remains 
significant scatter among recent experiments at low $Q^2$.

\subsubsection{Neutron charge radius}
\label{sec:neutron-radius}

The charge radius for the neutron can be expressed in terms of
Dirac and Pauli form factors as
\begin{equation} 
\langle r^2 \rangle_n 
= -6 \left. \frac{d F_{1n}(Q^2) }{dQ^2} \right|_{Q^2\rightarrow 0}
+\frac{3\kappa_n }{2m^2}
\end{equation}
where the first term is sometimes described as the intrinsic radius
({\it e.g.} \cite{Aleksandrov99,Glozman99})
while the second term is called the Foldy term and is
attributed to a charge separation induced by the zitterbewegung
motion of the magnetization density.
We prefer to describe the first term as the Dirac radius because it is
derived from the initial $Q^2$ dependence of the Dirac form factor 
and to reserve the term intrinsic radius for Eq. (\ref{eq:neutron_radius}),
which is based upon a moment of the radial function that we identified
as the intrinsic charge density. 
However, for the present purposes it will be clearer to refer to the radius
based upon the Sachs form factor as the Sachs radius.
The observation that the Foldy term, equal to -0.126 fm$^2$, 
is by itself almost equal to the mean-square charge radius obtained from 
$b_{ne}$ has generated considerable discussion.
Furthermore, as discussed in Sec.\ \ref{sec:data}, a substantial  
disagreement remains between the results from Dubna and those from
Oak Ridge, Garching, and Argonne.
For example, Alexandrov \cite{Aleksandrov99} argues that the Foldy 
term should be discarded and that the pionic cloud should
make the mean-square Dirac radius negative.
He further claims that experiments giving $b_{ne} \sim -1.31$ fm are
likely to suffer from serious experimental or interpretative errors 
because the corresponding mean-square Dirac radius would be positive.
On the other hand, Glozman and Riska \cite{Glozman99} calculated that the
pion loop contribution to the Dirac radius is negligible.  
Using a Foldy-Wouthuysen analysis of the interaction of a particle with
internal structure with an external electromagnetic field, 
Bawin and Coon \cite{Bawin99} demonstrated that the Foldy term
is cancelled by a higher-order term arising from the Dirac form factor, 
leaving the Sachs radius as the dominant coefficient,
independent of the dynamics responsible for the neutron form factors.

From a more microscopic point of view, 
Isgur used a quark model to argue that the observation of a very small 
Dirac radius is a potentially misleading accident and that the relativistic 
boost cancels the Foldy term such that the slope of $G_{En}$ does provide 
the second moment of the intrinsic charge distribution \cite{Isgur99}.
Cardarelli and Simula \cite{Cardarelli99,Cardarelli00} 
showed that this cancellation depends upon neglect of transverse 
momenta and that quark spin-spin interactions that break SU(6) symmetry
provide a mixed-symmetry $S^\prime$ component that enhances $G_{En}$ and 
provides a fairly accurate fit to the recoil-polarization data for 
$\mu_p G_{Ep}/G_{Mp}$. 
Leinweber {\it et al}.\ \cite{Leinweber01} argued that chiral perturbation
theory provides model-independent constraints on the dependencies of
nucleon magnetic moments and charge radii upon quark masses that 
demonstrate that the similiarity between the Sachs radius and
the Foldy term is purely accidental.

\begin{figure}[hbtp]
\centerline{ \strut\psfig{file=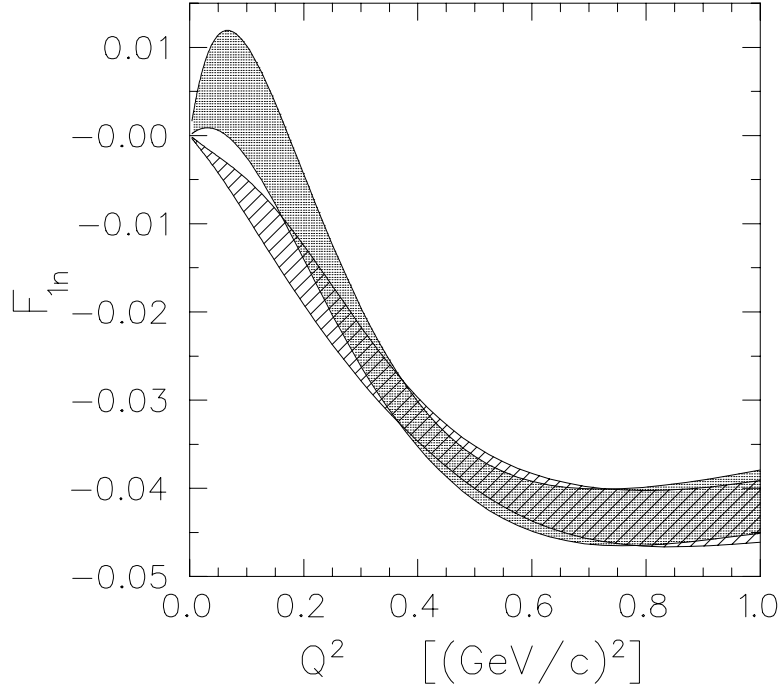,width=4.0in} }
\caption{Sensitivity to the Dirac radius:
the hatched (dotted) bands were obtained from fits to the $G_{En}$ data
that include (omit) the Oak Ridge datum for the neutron charge radius.}
\label{fig:F1n}
\end{figure}

Our fitted mean-square radius for the neutron charge density 
is largely determined by and is completely consistent with the 
datum of Ref. \cite{Kopecky97} that was included in the analysis.
This quantity is free of discrete ambiguity because $M_0$ vanishes
and the analysis procedure enforced that constraint explicitly.
Unfortunately, the electron-scattering data are not sufficiently 
precise at very low $Q^2$ to resolve the controversy concerning
the sign of the Dirac radius.
In the absence of a datum for $M_2$, the fit to the data for $G_{En}$
gives $M_2 = -0.187 \pm 0.04$ fm, which is about one or two standard 
deviations larger in absolute magnitude than the Dubna or Oak Ridge results, 
respectively, but is much less precise than either ---
the error bands on $G_{En}$ with or without the $M_2$ datum
overlap almost completely, even at low $Q^2$.
The effect of the constraint on $M_2$ is shown in 
Fig.\ \ref{fig:F1n},
where the shaded band for $F_{1n}$ represents the unconstrained fit 
to $G_{En}$ data that has a negative Dirac radius
while the cross-hatched band with a positive Dirac radius includes the 
$M_2$ datum and is consequently narrower as $Q^2 \rightarrow 0$.
The rather small difference between these fits is confined to 
$Q^2 \lesssim 0.2$ (GeV/$c$)$^2$, 
but because $F_{1n} < 0$ over most of its measured range
the fit with negative Dirac radius requires a sign change near 
$Q^2 \approx 0.14$ (GeV/$c$)$^2$.
Therefore, the present electron-scattering data offer very little
sensitivity to $M_2$. 
Furthermore, because the nuclear physics corrections needed to extract $G_{En}$
for low $Q^2$ are substantial even for polarization methods, it is unclear
whether one can ever expect better accuracy from nuclear physics than
atomic physics measurements of $M_2$.
We consider neither the theoretical argument for negative Dirac radius
nor the limited experimental evidence for a sign change in $F_{1n}$ at 
low $Q^2$ compelling.

\section{Discussion}
\label{sec:discussion}

\subsection{Comparison with Nonrelativistic Analyses}
\label{sec:nonrel}

\begin{figure}[hbtp]
\centerline{ \strut\psfig{file=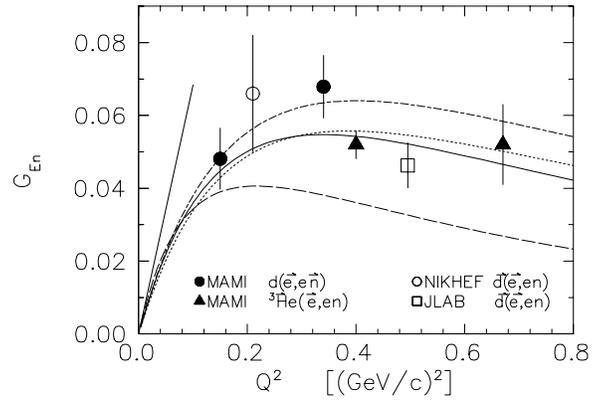,width=3.0in,angle=90} }
\caption{Selected data for $G_{En}$ at low $Q^2$ are compared with 
fits based upon the Galster model.
The solid curve is the original Galster fit while the dotted line is
a new fit based upon the entire data set considered in this work.
The dashed curve is a fit by 
Platchkov {\it et al}.\ {\protect\cite{Platchkov90}}
to elastic scattering
by deuterium based upon the Paris potential.
The dash-dotted curve was fitted to a subset of the Mainz data
by Schmieden {\protect\cite{Schmieden99}}.}
\label{fig:lowq}
\end{figure}

\begin{figure}[hbtp]
\centerline{ \strut\psfig{file=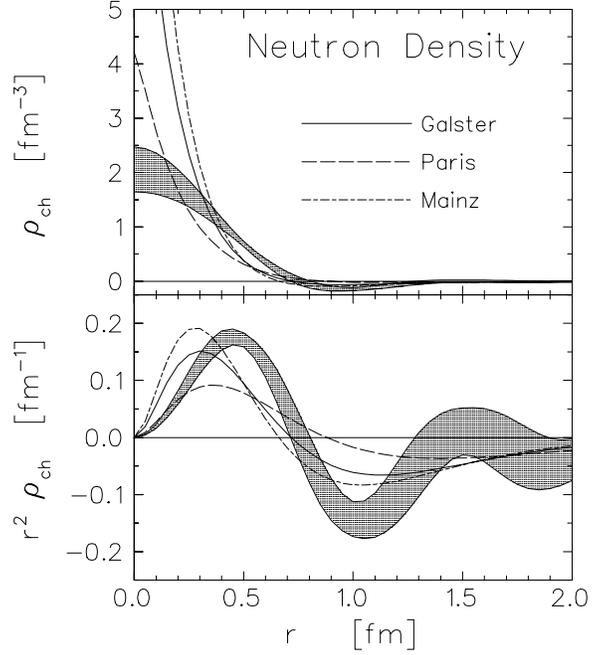,width=3.0in} }
\caption{Neutron charge densities obtained from nonrelativistic
Fourier transform of fits using the Galster model are compared with
the present results (band) using the relativistic transformation 
with $\lambda_E=0$.}
\label{fig:nonrel}
\end{figure}

Several analyses have appeared recently in which the parameters of the
Galster model were fitted to selected data for $G_{En}$ at low $Q^2$
and a density extracted using the nonrelativistic inversion formula 
given by Eq. (\ref{eq:NR}).
Examples of this type are shown in Figs. \ref{fig:lowq}-\ref{fig:nonrel}.
Figure \ref{fig:lowq} compares fits of this type to recent data using
the d$(\vec{e},e^\prime \vec{n})$, $\vec{d}(\vec{e},e^\prime n)$, or
$^3\vec{\rm He}(\vec{e},e^\prime n)$ reactions.
The slope of the form factor obtained from the Oak Ridge value for $b_{ne}$ 
is shown as a line segment.
The result obtained by Platchkov {\it et al.}\ \cite{Platchkov90}
from an analysis of elastic cross sections for electron scattering from 
deuterium using the Paris potential is shown as the dashed curve and 
lies well below the data obtained from polarization measurements.
However, variations of $\pm 50\%$ in $G_{En}$ were found using different
realistic nucleon-nucleon potentials; 
the result using the Paris potential is quoted most often, but the result
using the Argonne V14 potential is closer to the modern data.
The dash-dotted curve shows a fit by Schmieden \cite{Schmieden99}
to the data from Mainz, excluding the point for 
$^3\vec{\rm He}(\vec{e},e^\prime n)$ at $Q^2=0.4$ (GeV/$c$)$^2$; 
this fit gives the highest values for $G_{En}$.
The original Galster model is shown as a solid curve and our fit 
based upon Eq.\ (\ref{eq:Galster}) is shown as a dotted curve
and lies between the Paris and Mainz results.
The fit by Zhu {\it et al.}\ \cite{Zhu01} is also close to the dotted curve.
Note that our fit uses the entire data set described in Sec.\ \ref{sec:data}, 
including the slope at the origin, while the Mainz fit used a smaller subset.
Even though the data employed by Galster {\it et al.}\ \cite{Galster71}
had much larger uncertainties, 
their result is remarkably close to the present analysis.  
However, the apparent agreement of the original Galster parametrization 
with more modern data must be judged as fortuitous.

Buchmann {\it et al}.\ \cite{Buchmann97a,Grabmayr01} argue that the 
neutron charge density reflects differences between the spatial
distributions of constituent quarks induced by the color hyperfine
interaction and that because the same interaction is responsible
for the $N-\Delta$ mass splitting the $N \rightarrow \Delta$
quadrupole form factor is related in a simple manner to $G_{En}$.
In Ref.\ \cite{Grabmayr01} they used the Galster parametrization
to fit a selection of the $G_{En}$ data and used nonrelativistic inversion
to obtain a density similar to the solid curve in Fig. \ref{fig:nonrel}.

The nonrelativistic densities obtained from these fits are compared
in Fig. \ref{fig:nonrel} to the present relativistic LGE result.
Here we chose $\lambda_E=0$ to minimize the differences between
relativisitic and nonrelativistic inversion formulas; 
however, it is important to remember that the Galster form factor 
is inconsistent with the relativistic inversion procedure for 
$\lambda_E=0$.
The upper panel, emphasizing the interior density, 
demonstrates that the naive Fourier transform tends to produce a rather 
hard core and an unphysical cusp at the origin.
The relativistic transformation, by contrast, softens the interior
density and eliminates its cusp, providing a much more plausible
charge density.
The lower panel, emphasizing the surface lobe, shows that all three
nonrelativistic densities have positive peaks at smaller radii
and less surface charge than our result.

\subsection{Approach to scaling}
\label{sec:scaling}

\begin{figure}[hbtp]
\centerline{ \strut\psfig{file=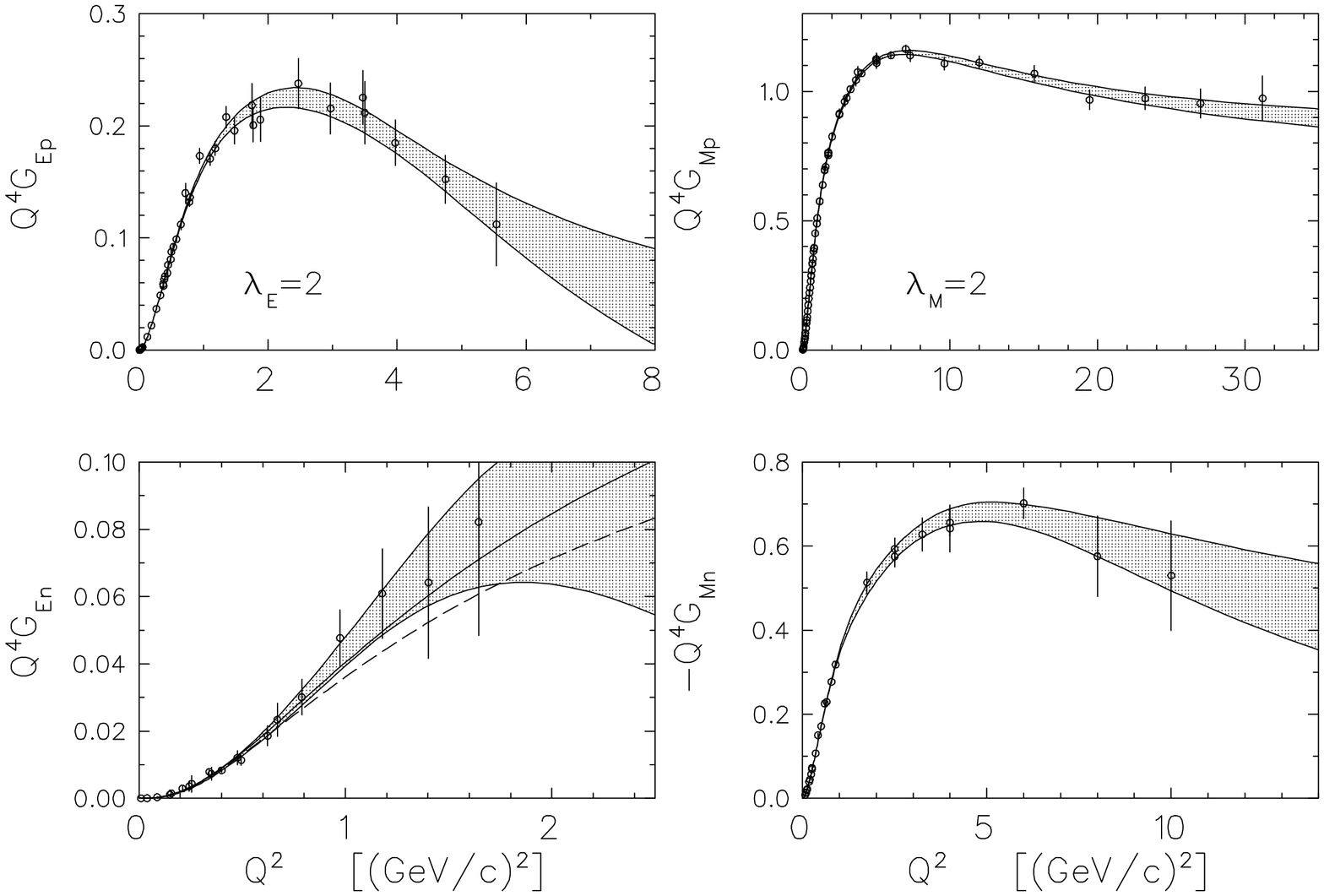,width=6.5in,angle=90} }
\caption{The approach to scaling is shown by multiplying Sachs form
factors by $Q^4$.
For $G_{En}$ the original Galster model is shown as a dashed curve and
a new fit as the solid curve.
The LGE parametrization was used with $\lambda_E=\lambda_M=2$.}
\label{fig:pQCD}
\end{figure}

The asymptotic behavior of the fitted form factors is illustrated in
Fig. \ref{fig:pQCD} by multiplication by $Q^4$.
The uncertainties in the fitted form factors are clearly dominated by
experimental uncertainties where data are available, while the expansion
of the error bands for larger $Q^2$ is governed by the incompleteness
errors in the range $k_{max}<k<k_m$.
The scaling behavior of $G_{Mp}$ appears fully developed because the
data up to $Q_{max}^2 = 31$ (GeV/$c$)$^2$ reach a sufficiently 
large ratio $k_{max}/k_m = 0.95$ to strongly constrain the 
asymptotic limit.
Although the fit to $G_{Mn}$ is compatible with pQCD scaling, the
uncertainties at large $Q^2$ grow much more rapidly because here
$k_{max}/k_m = 0.86$ is smaller and the data for $k_{max}/k_m > 0.79$ 
are much less precise.
Furthermore, because scaling is not fully developed for $G_{Mp}$
until $Q^2 \gtrsim 20$ (GeV/$c$)$^2$ we should expect that data for $G_{Mn}$ 
at higher $Q^2$ will be needed to establish its asymptotic limit. 
The data for $G_{Ep}$ do not show scaling behavior for
$Q_{max}^2 = 5.54$ (GeV/$c$)$^2$, where $k_{max}/k_m = 0.78$ ---
despite relatively large uncertainties in extrapolation, 
the present fit suggests a sign change near $Q^2 \sim 10$ (GeV/$c$)$^2$.
The new recoil polarization data suggest that $G_{Ep}$ and $G_{Mp}$ 
differ dramatically for $Q^2 \gtrsim 3$ (GeV/$c$)$^2$ and it is
clearly crucial to extend the $G_{Ep}$ data to higher $Q^2$.
It will also be important to check those data using another technique.
The results of an improved Rosenbluth experiment are expected 
soon \cite{JLAB01-001}.

The present data for $G_{En}$ are too limited in both range and precision 
to address the question of scaling.
High-precision data for $Q^2 \lesssim 1.5$  (GeV/$c$)$^2$ from
a recent d$(\vec{e},e^\prime \vec{n})$ experiment are expected soon
\cite{TJNAF93-038},
and an approved proposal for $^3\vec{\rm He}(\vec{e},e^\prime n)$
should extend the range to 3.4 (GeV/$c$)$^2$ in a couple of years
\cite{JLAB02-013}.
However, experience with the other three form factors suggests that
one must approach 20 (GeV/$c$)$^2$ to determine the asymptotic limit.

\begin{figure}[hbtp]
\centerline{ \strut\psfig{file=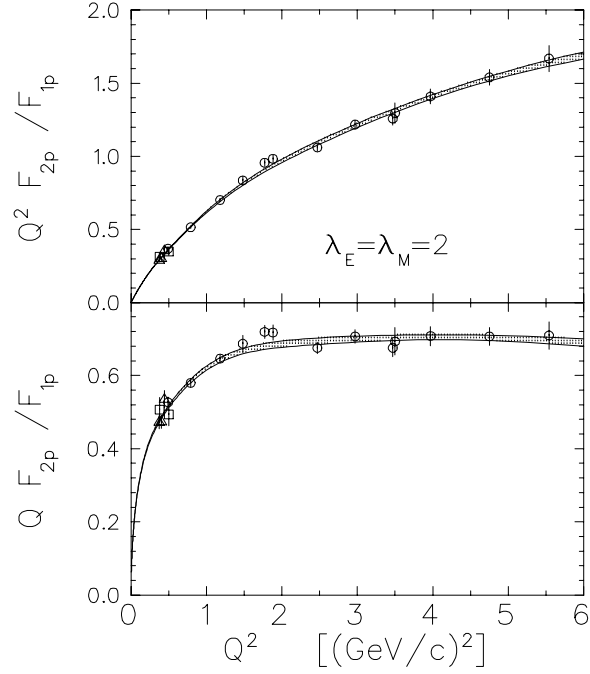,width=3.0in,angle=90} }
\caption{Data for $F_2/F_1$ are compared with the usual $Q^{-2}$
scaling expected from pQCD or with $Q^{-1}$ scaling recently
proposed by several authors.
Fitted bands employ the LGE parametrization with $\lambda_E=\lambda_M=2$.
The data are shown as  
squares {\protect\cite{Milbrath99}}, 
triangles {\protect\cite{Pospischil01}},
circles {\protect\cite{MKJones00,Gayou02}}.
}
\label{fig:f2f1p_low}
\end{figure}

Quark helicity conservation suggests that $Q^2 F_2/F_1$ should approach
a constant in the asymptotic limit of large $Q^2$.
The Dirac and Pauli form factors are related to Sachs form factors by
\begin{mathletters}
\begin{eqnarray}
F_1 &=& \frac{G_E + \tau G_M}{1+\tau} \\
\kappa F_2 &=& \frac{G_M - G_E}{1+\tau}
\end{eqnarray}
\end{mathletters}
and their ratio is given by
\begin{equation}
\frac{\kappa F_2}{F_1} = \frac{1-g}{\tau+g}
\end{equation}
where $g=G_E/G_M$ can be measured directly by either recoil or target
polarization.
Recoil-polarization data for this ratio are compared in 
Fig.\ \ref{fig:f2f1p_low}
with bands constructed from the present fits to Sachs form factors.
Gayou {\it et al.}\ \cite{Gayou02} observed that the proton recoil 
polarization data appear to reach a plateau 
in the range $2 < Q^2 <6$ (GeV/$c$)$^2$ when scaled by $Q$ instead 
of the expected $Q^2$.
Ralston {\it et al.}\ \cite{Ralston00,Ralston02}
suggested that orbital angular momentum in the quark 
distribution could explain an asymptotic behavior of the form
$F_2/F_1 \propto Q^{-1}$.
Later Miller and Frank \cite{Miller02}
argued that substantial violation of quark helicity 
conservation should be expected for intermediate $Q^2$ when Poincar\'e 
invariance is imposed upon relativistic constituent quark models.

\begin{figure}[hbtp]
\centerline{ \strut\psfig{file=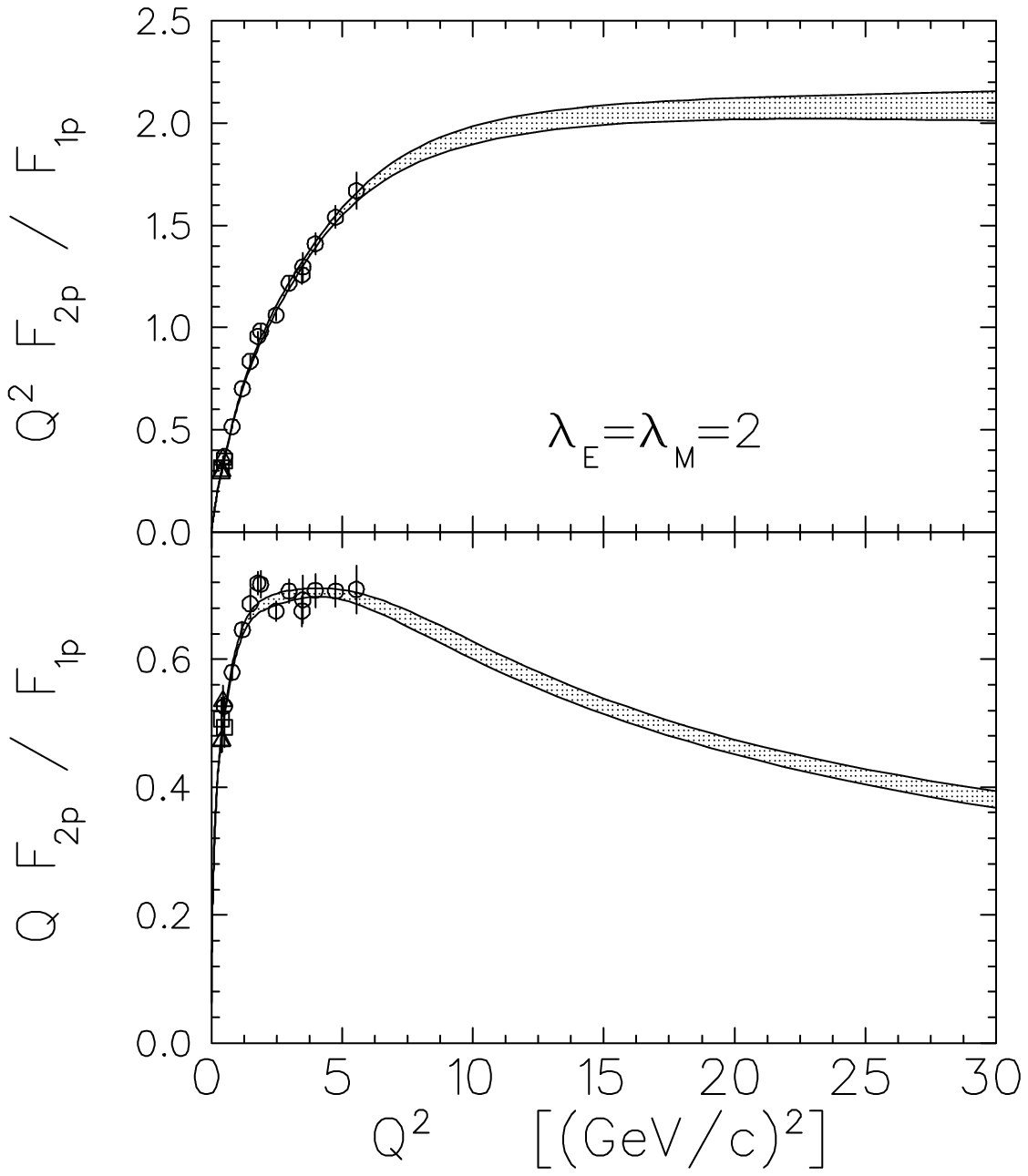,width=3.0in,angle=90} }
\caption{Data for $F_2/F_1$ are compared with the usual $Q^{-2}$
scaling expected from pQCD or with $Q^{-1}$ scaling recently
proposed by several authors.
Fitted bands employ the LGE parametrization with $\lambda_E=\lambda_M=2$.}
\label{fig:f2f1p}
\end{figure}

Recognizing that $g$ is small compared with $\tau$ for 
$Q^2 > 6$ (GeV/$c$)$^2$ and that the model imposes constraints upon 
the Sachs form factors for high $Q^2$ that inhibit the growth of the 
error band for $F_2/F_1$, we can extrapolate $F_2/F_1$ beyond the
present experimental range for $G_{Ep}$.
This extrapolation, shown in Fig.\ \ref{fig:f2f1p}, shows that the
data are consistent with quark helicity conservation for 
$Q^2 \gtrsim 20$ (GeV/$c$)$^2$.
Therefore, although the present model is consistent with the observation 
by Gayou {\it et al}.\ that $Q F_2/F_1$ is approximately constant 
in the range $2 < Q^2 < 6$ (GeV/$c$)$^2$, we attribute that observation 
to the appearance of a broad maximum in $Q F_2/F_1$ rather than to the 
onset of true $Q^{-1}$ scaling.

\subsection{Quark densities}
\label{sec:quarks}

Assuming isospin symmetry and neglecting strange quarks,
nucleon charge densities can be expressed in a simple two-flavor 
quark model as
\begin{mathletters}
\begin{eqnarray}
\rho_p(r) &=& \frac{4}{3} u(r) - \frac{1}{3} d(r) \\
\rho_n(r) &=& - \frac{2}{3} u(r) + \frac{2}{3} d(r)
\end{eqnarray}
\end{mathletters}
where $u(r)$ is the radial distribution for an up quark in the 
proton or a down quark in the neutron while $d(r)$ is the distribution 
for a down quark in the proton or an up quark in the neutron.
Thus, the quark densities are obtained from nucleon charge densities using
\begin{mathletters}
\begin{eqnarray}
u(r) &=& \rho_p(r) + \frac{1}{2} \rho_n(r) \\
d(r) &=& \rho_p(r) + 2 \rho_n(r)
\end{eqnarray}
\end{mathletters}
where $u(r)$ and $d(r)$ are normalized to unity according to 
\begin{equation}
\int_0^\infty dr \; r^2 q(r) = 1 \\
\end{equation}
but need not be positive everywhere.
There is no guarantee that these combinations of radial densities obtained
from form factor data by relativistic inversion must be positive, nor are 
the densities derived from positive-definite matrix elements.
Within the quark model one could decompose the densities
\begin{equation}
\label{eq:quarks}
q(r) = q_v(r) + q_s(r) - \bar{q}_s(r)
\end{equation}
for each flavor, $q$, into valence ($v$) and sea ($s$) contributions
which would be expected to be positive definite 
but cannot be separated using only Sachs form factors.
The valence distributions, $q_v$, are normalized to unity while
the sea distributions, $q_s$ and $\bar{q}_s$, must have equal
normalizations.
Thus, within this picture the contributions from quarks should be positive 
but antiquarks in the sea could produce regions where $u(r)$ or $d(r)$ 
might become negative. 

\begin{figure}[hbtp]
\centerline{ \strut\psfig{file=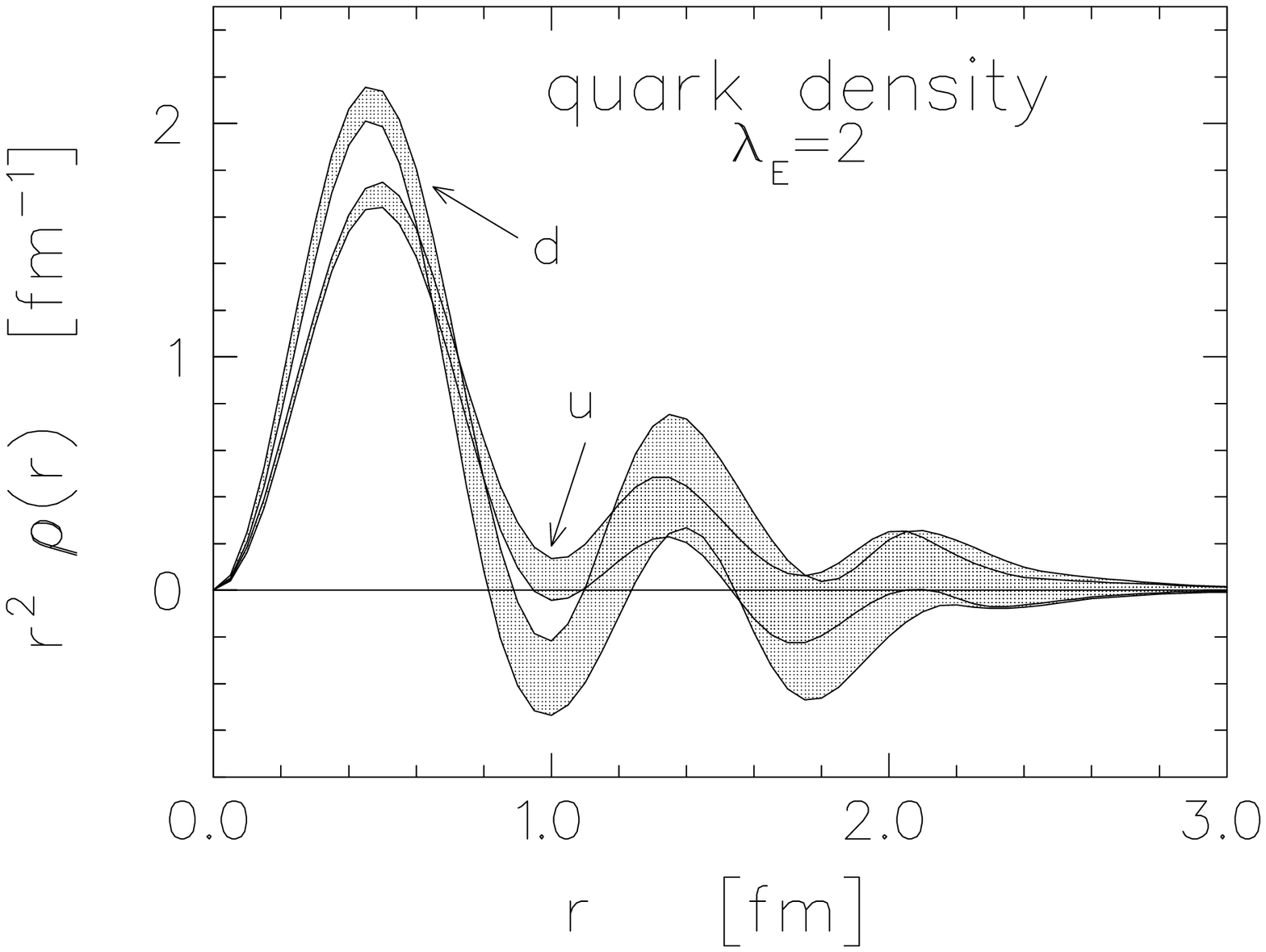,width=3.0in,angle=90} }
\caption{Quark densities obtained from proton and neutron charge densities 
using the LGE parametrization with $\lambda_E=2$.}
\label{fig:quarks2_lge2}
\end{figure}

The quark densities obtained with LGE densities for $\lambda_E=2$ that
are displayed in Fig.\ \ref{fig:quarks2_lge2} are determined with
relatively small uncertainties and are predominantly positive, as expected.
We find that the pair of like quarks has a somewhat broader distribution than
that of the unlike quark, being depleted in the interior and enhanced at
the surface.
In this model, the neutron charge density arises from incomplete cancellation
between charge densities for up and down quarks, resulting in positive
core and negative surface charges.
The broader distribution for like quarks is consistent with the repulsive
color hyperfine interaction between like quarks needed to explain the 
$N-\Delta$ mass difference.
This picture is also consistent with the model of a pion cloud surrounding
a three-quark core, 
but in that model one might expect to find a slightly negative $d(r)$ 
near the surface due to the antiquark content of the pion.
The present data are marginally consistent with a slightly negative $d(r)$
near 1.0 fm, but more accurate data for the neutron charge density would
be needed to reduce the uncertainty in $d(r)$ before drawing a definitive 
conclusion.

\subsection{Discrete Ambiguities}
\label{sec:ambiguities}

\begin{figure}[ht]
\centerline{ \strut\psfig{file=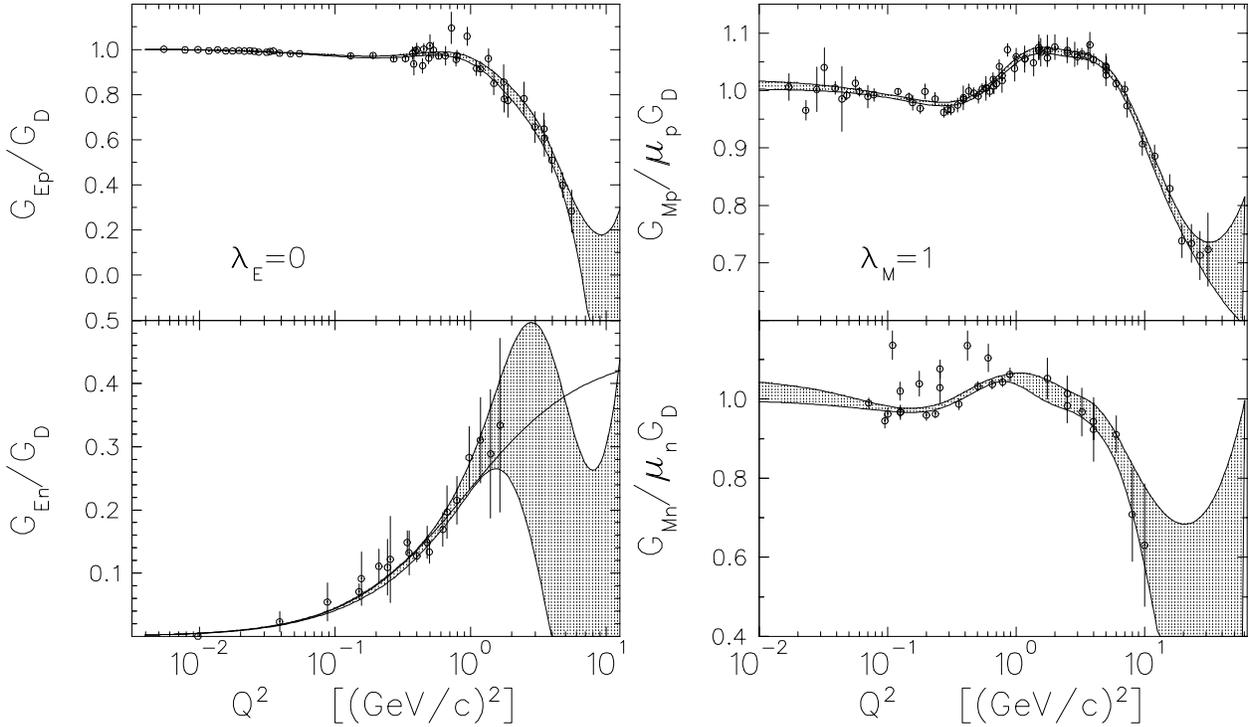,width=6.5in,angle=90} }
\caption{The bands show fits to selected data for nucleon electromagnetic 
form factors using the LGE parametrization with $\lambda_E=0$ and 
$\lambda_M=1$ as suggested by the soliton model.}
\label{fig:Gsoliton}
\end{figure}

Although discrete ambiguities in $\lambda_E$ and $\lambda_M$ do not affect
fitted form factors in the range where data are available, 
the choice of $\lambda$ does affect the growth of the uncertainties
in extrapolated form factors as $Q^2$ increases beyond the measured range.
For example, Fig.\ \ref{fig:Gsoliton} shows form factors fitted using the
LGE parametrization with $\lambda_E=0$ and $\lambda_M=1$ as
suggested by the relativistic soliton model.
Although the relative uncertainties become quite large for 
$Q^2$ beyond the range of the experimental data,
the uncertainties in the form factors actually remain small
because, with the exception of $G_{En}$, the form factors are greatly
reduced as their experimental limits are reached.
Consequently, the contribution of the uncertainties in form factors at large
$Q^2$ upon uncertainties in radial densities is relatively insensitive to
the choice of $\lambda$.
However, the change in fitted density due to a change of $\lambda$ need not be
contained within the fitted error band --- 
the bands do not accommodate discrete ambiguities in the model. 

\begin{figure}[hbtp]
\centerline{ \strut\psfig{file=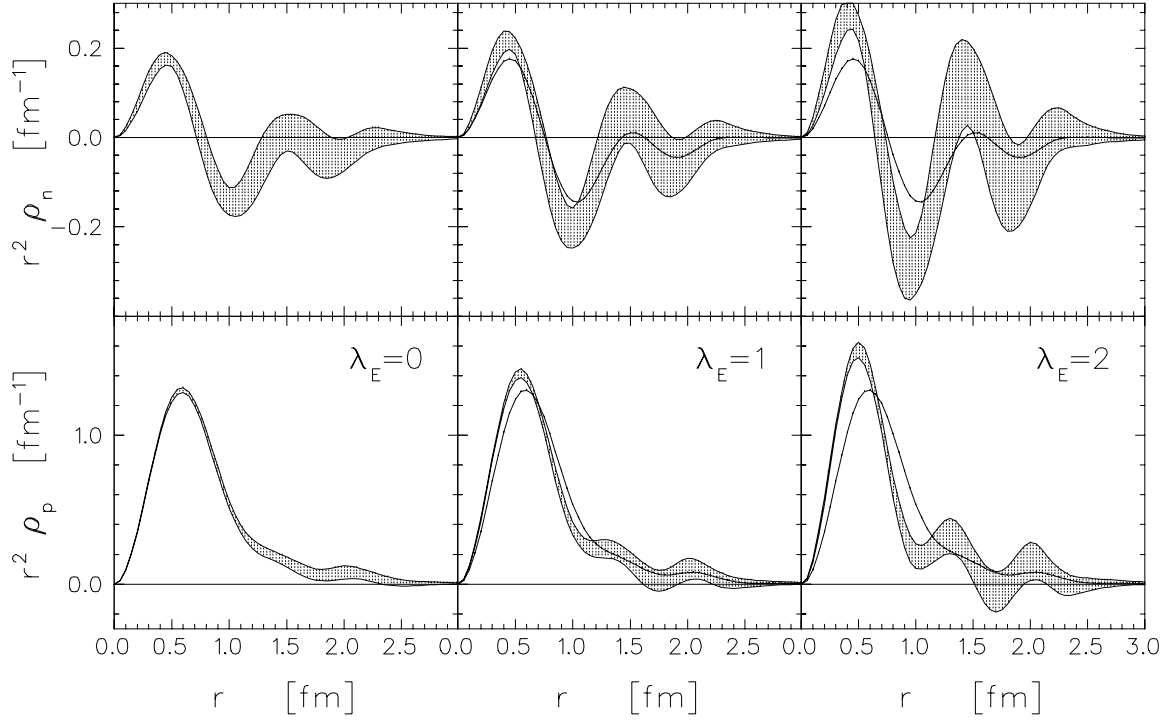,width=6.0in,angle=90} }
\caption{Discrete ambiguities in the charge densities.  
The value of $\lambda_E$ increases from 0 to 2 from left to right.
The curves for $\lambda_E=0$ are reproduced, without error bands, in
the middle and right columns for comparison.}
\label{fig:ambiguities}
\end{figure}

The sensitivity of fitted charge densities to the choice of $\lambda_E$
is illustrated in Fig. \ref{fig:ambiguities}.
These figures were made with the LGE parametrization, but very similar 
results are obtained with the FBE parameterization.
The smoothest results at large radii are obtained with $\lambda=0$, 
whereas larger values of $\lambda$ tend to pull the density inward and 
to amplify oscillations at large radii.
This behavior can be understood by interpreting Eq. (\ref{eq:rhok})
in terms of the convolution theorem for Fourier transforms.
Expressing this equation in terms of $k$, 
\begin{equation}
\tilde{\rho}(k) = (1-\frac{k^2}{k_m^2})^{-\lambda}
G(\frac{k^2}{1-\frac{k^2}{k_m^2}})
\end{equation}
one finds that $\tilde{\rho}(k)$ for $\lambda>0$ is obtained from the Lorentz
contracted form factor by deconvolution of a resolution function with
mean square radius equal to $3\lambda/2m^2$.
This resolution function originates in the zitterbewegung and is 
characterized by the nucleon Compton wavelength.
As discussed in Sec.\ \ref{sec:results-densities}, acceptable fits to
the Sachs form factor data for $Q^2 \sim 1$ (GeV/$c$)$^2$ using $\lambda=2$
seem to require structure in the radial densities in the 1-1.5 fm region.
Reducing $\lambda$ tends to smooth out such structures, 
but sacrifices the high $Q^2$ limit. 
Therefore, although $\lambda=2$ provides the most natural implementation of 
pQCD scaling, accurate reproduction of the data using that representation 
of the intrinsic form factor appears to require oscillations in the 
radial density with a wavelength of order 0.7 fm. 

\begin{figure}[hbtp]
\centerline{ \strut\psfig{file=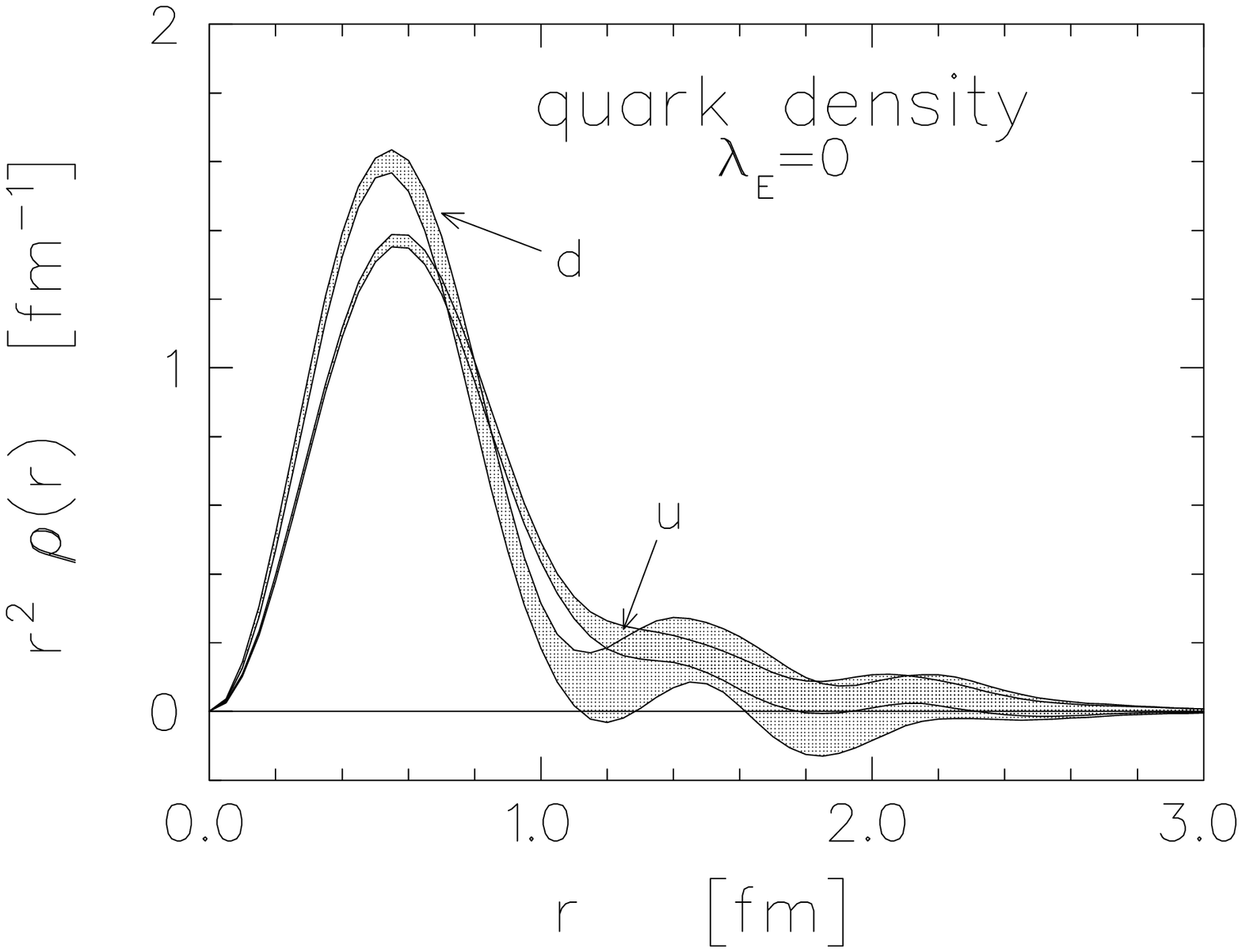,width=3.0in,angle=90} }
\caption{Quark densities obtained from proton and neutron charge densities 
using the LGE parametrization with $\lambda_E=0$.}
\label{fig:quarks2_lge0}
\end{figure}

Despite this ambiguity in the relationship between form factor and density, 
the qualitative conclusion that $u(r)$ is broader than $d(r)$ depends only 
the assumption of isospin symmetry and the observation that the neutron 
charge density is positive in the interior and negative at its surface.
The quark distributions derived using $\lambda_E=0$ shown in 
Fig.\ \ref{fig:quarks2_lge0}
are qualitatively similar to those shown for $\lambda_E=2$ in
Fig.\ \ref{fig:quarks2_lge2},
but are slightly more diffuse.
The choice $\lambda_E=0$ is appropriate for the soliton model, 
is consistent with nonrelativistic expectations for small $Q^2$, 
and is favored by the radius obtained from the Lamb shift.
However, it appears to be inconsistent with pQCD.
Therefore, in the absence of a unique relationship between form factors 
and densities, it appears necessary to select the appropriate value of
$\lambda$ based upon the intended application.
For long-wavelength properties one should use $\lambda_E=0$, 
but for extrapolation to the pQCD limit one should employ 
$\lambda_E=\lambda_M=2$.

\subsection{Alternative parametrizations}
\label{sec:alternatives}

Without a unique relationship between form factor and intrinsic density, 
one may question the value of densities extracted by the present techniques.
It is clear that simple parametrizations like the dipole model cannot
fit the proton or $G_{Mn}$ data over a wide range of $Q^2$;
nor is it likely that the Galster parametrization will continue to
fit $G_{En}$ at higher $Q^2$.
The empirical parametrization proposed by Bosted \cite{Bosted95}
\begin{equation}
G \propto (1 + a_1 Q + a_2 Q^2 + a_3 Q^3 + a_4 Q^4)^{-1}
\end{equation}
fits the data for large $Q^2$ well and is consistent with pQCD, 
but its odd powers of $Q$ are incompatible with the interpretation 
of the form factor as the Fourier transform of a radial density and
with the moment expansion for small $Q^2$.
Furthermore, it is not sufficiently flexible to provide realistic error bands,
especially if the odd powers are eliminated.
By inclusion of a $Q^5$ term, Brash {\it et al.}\ \cite{Brash02} 
also sacrificed the pQCD limit in order to improve the quality of the 
fit for finite $Q^2$.

\begin{figure}[hbtp]
\centerline{ \strut\psfig{file=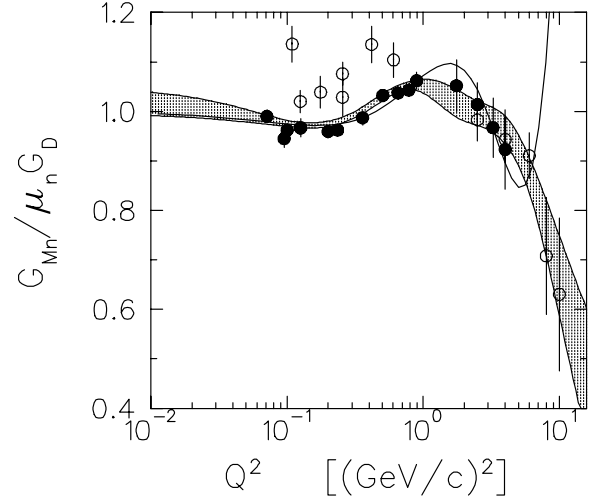,width=3.0in,angle=90} }
\caption{Comparison with the $G_{Mn}$ analysis by Kubon {\it et al.}
{\protect\cite{Kubon02}}.  
The band shows our LGE fit with $\lambda_M=2$ to the entire data
set while the solid line shows the continued-fraction fit by
Kubon {\it et al.} {\protect\cite{Kubon02}} to a subset of the data.
The data included (omitted) by Kubon {\it et al}.\ are indicated by
filled (open) circles.}
\label{fig:kubon}
\end{figure}

Kubon {\it et al.} \cite{Kubon02} fit a subset of the $G_{Mn}$ data using
a continued-fraction parametrization of the form
\begin{equation}
G_{Mn}(Q^2) = \frac{\mu_n}{1+
\frac{b_1 Q^2}{1+\frac{b_2 Q^2}{1+...}}}
\end{equation}
carried to fifth order.
This parametrization provides a good fit to the data for 
$Q^2 < 4$ (GeV/$c$)$^2$ using 5 parameters, 
but the parameters do not decrease with order and the fit depends
upon fairly delicate cancellations.
Adding additional terms to extend the range of $Q^2$ changes the 
lower terms.
Furthermore, this fit does not conform with the asymptotic $Q^{-4}$ 
behavior expected by pQCD unless a fairly complicated constraint 
of the form 
\begin{equation}
b_3 b_5 + b_2(b_4 + b_5) = 0
\end{equation}
is imposed to eliminate the $Q^{-2}$ contribution.
The constraint increases in complexity as additional terms are included.
A comparison between the Kubon parametrization and our LGE fit with
$\lambda_M=2$ is shown in Fig.\ \ref{fig:kubon}.
The Kubon analysis included only the data indicated by filled circles
and was limited to $Q^2 < 4$ (GeV/$c$)$^2$ while our analysis used
all data shown and extended to 10 (GeV/$c$)$^2$.
The LGE parametrization fits the data well over a broader range and is 
compatible with pQCD,
whereas the continued-fraction parametrization behaves badly soon after 
the range fitted by Kubon {\it et al}.
By not imposing the pQCD constraint, the extrapolation deteriorates quite 
quickly. 
We believe that the continued-fraction method also underestimates the
uncertainty in the rms radius due to the strong correlations among
its parameters and its built-in normalization constraint at $Q^2=0$.

A rather different phenomenological parametrization can be made in
the context of the vector meson dominance model at modest $Q^2$ 
matched to pQCD at large $Q^2$, denoted VMD+pQCD.
That approach was pioneered by Gari and Kr\"umpelmann
\cite{Gari85,Gari86,Gari92a,Gari92b} 
and recently refined by Lomon \cite{Lomon01,Lomon02}.
Similarly, the classic dispersion-theory analysis of
H\"ohler {\it et al}.\ \cite{Hohler76} has recently been updated 
by Mergell {\it et al}.\ \cite{MMD} to handle better the 
requirements of unitarity and the approach to the pQCD limit.
These approaches have the advantage that all four electromagnetic
form factors are analyzed simultaneously, thereby relating their
isospin structure to an underlying model, and can be extended 
to timelike momentum transfer \cite{HMD}. 
By contrast, our approach is limited to spacelike momentum transfer
and must construct the isospin form factors from four independent 
fits to individual form factors.
Both the VMD and dispersion theory approaches appear to be capable of 
fitting the data as well as our linear expansion analysis, although
the data have improved considerably since the analysis of 
Ref. \cite{MMD}.
However, we omit detailed comparisons here because these models
do not consider radial densities.

\subsection{Importance of $G_{En}$ data at higher $Q^2$}
\label{sec:highQ2}

\begin{figure}[hbtp]
\centerline{ \strut\psfig{file=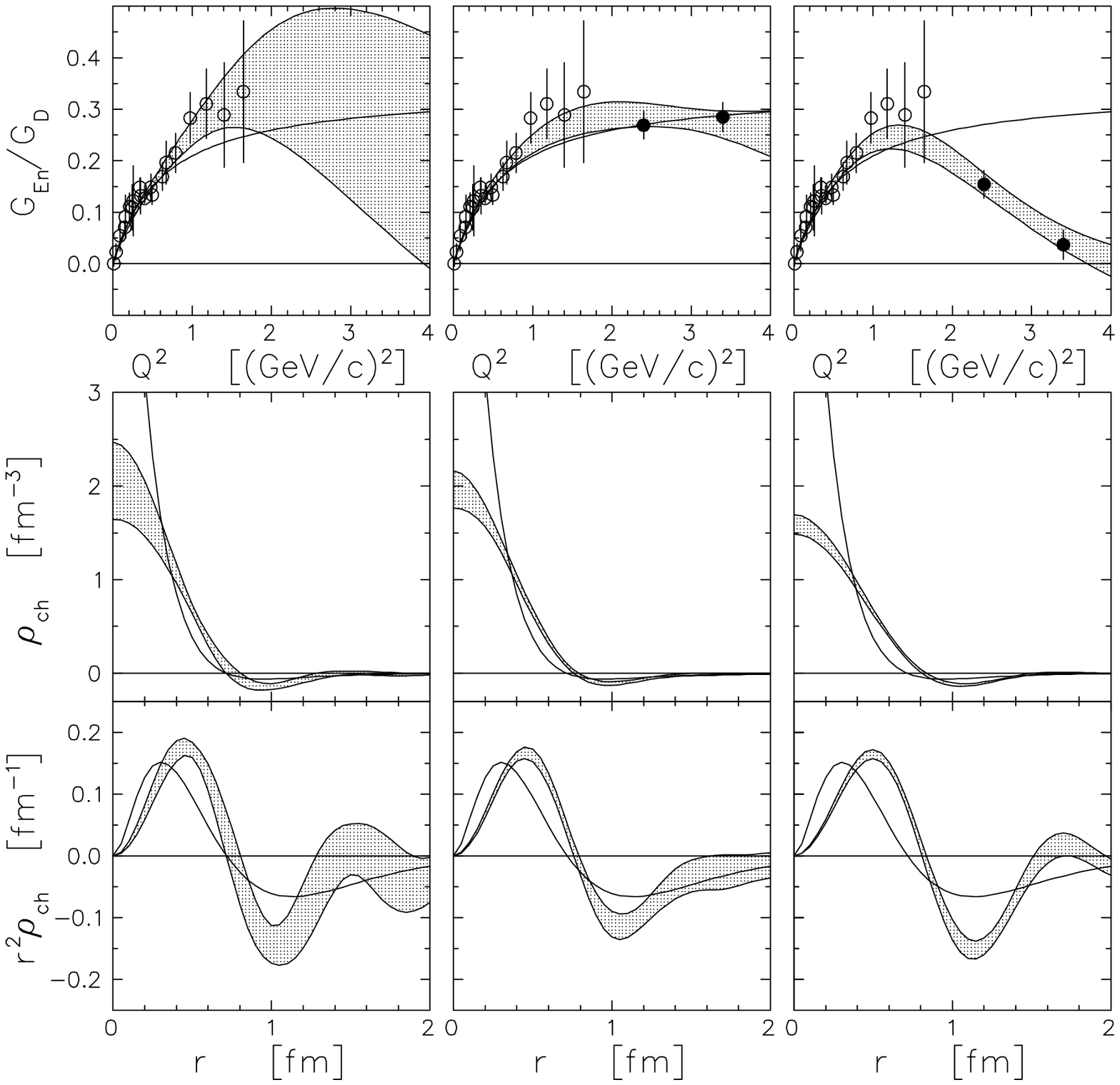,width=5.0in} }
\caption{The sensitivity of the neutron charge density to extension of
the experimental range of $Q^2$ is illustrated by comparing a fit to
published data (open circles) with two scenarios that include 
pseudodata (filled circles) for $Q^2 = 2.4, 3.4$ (GeV/$c$)$^2$.
The middle column assumes that new data would follow the Galster
parametrization, shown by the solid curve, 
while the right column assumes that $G_{En}/G_D$ would decrease 
for $Q^2 > 2$ (GeV/$c$)$^2$.
Densities for the Galster model use nonrelativistic inversion.
All fits use the LGE parametrization and relativistic inversion
with $\lambda_E=0$.}
\label{fig:highQ2}
\end{figure}

The present data for $G_{En}$ do not extend high enough in $Q^2$
to determine the interior charge density as accurately for the neutron 
as for the proton or to permit reliable extrapolation to the 
scaling regime, but new data expected from an approved proposal
\cite{JLAB02-013}
at Jefferson Laboratory should help considerably.
The impact of extending the $Q^2$ range to 3.4 (GeV/$c$)$^2$ is
illustrated in Fig.\ \ref{fig:highQ2}.
This analysis was performed using $\lambda_E=0$, which permits the
greatest latitude at high $Q^2$.
The left column shows the form factor and density fitted to 
published data (shown by open symbols), 
while the middle and right columns show the effect of pseudodata 
(shown as filled symbols)
for two hypothetical scenarios.
The middle scenario assumes that the new data would follow the Galster
parametrization while the right scenario assumes that the new
data would fall more rapidly than the dipole form factor for 
$Q^2 > 2$  (GeV/$c$)$^2$.
Both scenarios are compatible with the uncertainties extrapolated
from the fit to the present data and can be fitted well, with error
bands for $Q^2 \lesssim 4$  (GeV/$c$)$^2$ consistent with the
anticipated experimental precision.
The reduction of the incompleteness error obtained by extending
the measurements to higher $Q^2$ greatly improves the precision of
the interior charge density.
If measurements at higher $Q^2$ come close to the Galster parametrization,
then the error band would be reduced in width with little change in its
centroid.
On the other hand, if new measurements of $G_{En}/G_D$ decrease with $Q^2$ 
in a manner similar to the proton charge form factor, the softer charge
density would be reduced in the interior, moving toward the lower edge of 
the present error band. 
These scenarios have quite different asymptotic values for $Q^4 G_{En}$,
but the present data cannot distinguish between them.
Furthermore, the recent VMD+pQCD analysis by Lomon \cite{Lomon02}
suggests that $G_{En}/G_D$ could reach an asymptotic value substantially
higher than predicted by the Galster parametrization.
Therefore, it is very important to extend $G_{En}$ data as far as
possible in $Q^2$.

\section{Summary and Conclusions}
\label{sec:conclusions}

We have employed expansions in complete sets of radial basis functions
to parametrize nucleon Sachs form factors in terms of charge
and magnetization densities.
Our selection of data emphasizes recent polarization data.
The inversion from form factor to density is based upon relativistic
models in which the spatial frequency 
$k = Q/\sqrt{1+\tau}$
in the rest frame is related to the momentum transfer $Q$ in the Breit 
frame by Lorentz contraction.
The maximum possible frequency sampled by electron scattering is then 
$k<k_m$ where $k_m = 2m$ is determined by the nucleon Compton wavelength.
A variety of models produce inversion formulas of the form
\begin{displaymath}
\tilde{\rho}(k) = G(Q^2) (1+\tau)^\lambda
\end{displaymath}
but differ in the choice of $\lambda$.
By considering the asymptotic form of $G(Q^2)$, we can limit the exponent
$\lambda$ to 0,1,2.
The relativistic soliton model suggests $\{\lambda_E=0,\lambda_M=1\}$,
the original quark cluster model suggested $\lambda_E=\lambda_M=1$, and
a more symmetric version of the quark cluster model gives choices 
$\lambda_E=\lambda_M=2$ that are compatible with pQCD without the
somewhat artificial constraints upon $\tilde{\rho}(k_m)$ needed by
the other models.
In most of this paper we parametrized the Sachs form factors using
the Laguerre-Gaussian expansion (LGE) and derived densities using
$\lambda_E=\lambda_M=2$, but we have also analyzed the impact of the
discrete ambiguity in $\lambda$ upon the radial densities.
Although some of the details of the radial densities are affected
by the discrete ambiguity in the relativistic inversion formula,
their qualitative features are independent of $\lambda$.

We find that virtually identical fits to the Sachs form factors are
obtained with either LGE or Fourier-Bessel expansions (FBE) and within a
wide range these fits are independent of details of the parametrization, 
such as number of terms, radial scale parameters, and tail bias.
The fitting procedure uses $\tilde{\rho}(k)$ pseudodata for $k>k_{max}$ 
to estimate the incompleteness error in radial density due to the limitation 
of experimental data for the range $k<k_{max}<k_m$.
For a given choice of $\lambda$, the radial densities fitted with either
LGE or FBE expansions are practically identical and are insensitive
to details of the analysis.
We find that the proton charge density is significantly broader than its
magnetization density, consistent with the observation in recent 
recoil-polarization measurements that $G_{Ep}/G_{Mp}$ decreases in
an almost linear fashion for $1<Q^2<6$ (GeV/$c$)$^2$.
Our result for the proton Sachs charge radius is consistent with
a recent determination based upon the 1S Lamb shift in hydrogen.
Similarly, we find the magnetization density is slightly broader for the
neutron than for the proton.
Each of these three densities exhibits a secondary peak in $r^2 \rho(r)$
near $1-1.5$ fm that cannot be suppressed without sacrificing the fit
to the corresponding form factor for $Q^2 > 1$ (GeV/$c$)$^2$.
This structure appears somewhat stronger for $\lambda=2$ than for
$\lambda=0$, but $\lambda=2$ provides the clearest extrapolation to 
the pQCD limit. 

The recent recoil-polarization data for $F_{2p}/F_{1p}$ appear to favor
scaling with $Q^{-1}$ rather than the $Q^{-2}$ expected from quark
helicity conservation in pQCD.
Although that observation has stimulated some speculation about violations
of quark helicity conservation due to orbital angular momentum or
imposition of Poincar\'e invariance, 
in our analysis with $\lambda_E=\lambda_M=2$ we find that although
$Q F_{2p}/F_{1p}$ appears to be nearly constant for $2<Q^2<6$ (GeV/$c$)$^2$
we nevertheless obtain a constant asymptotic value for $Q^2 F_{2p}/F_{1p}$
for $Q^2 \gtrsim 20$ (GeV/$c$)$^2$ where $G_{Mp}$ scales with $Q^{-4}$.
Therefore, we find that the data are consistent with a broad maximum in 
$Q F_{2p}/F_{1p}$ and do not require true $Q^{-1}$ scaling.

We have compared the LGE parametrization for $G_{En}$ to fits based 
upon the Galster parametrization.  
Although the range of $Q^2$ remains too small to discriminate between
these models, the Galster parametrization cannot be inverted using a
relativistic relationship between intrinsic spatial frequency and
Breit-frame momentum transfer unless $\lambda \ge 1$.
The traditional nonrelativistic inversion of the Galster form factor
produces a charge density with an unphysical cusp at the origin
while the relativistic fit to the data using the LGE form factor softens 
the interior density and removes the cusp.
However, the incompleteness error in the neutron charge density remains
fairly large because the available $G_{En}$ data are limited to small $Q^2$ 
and the data for $0.4 < Q^2 < 1.6$ (GeV/$c$)$^2$ have relatively
large uncertainties.
More precise data for $0.5 < Q^2 < 1.5$ (GeV/$c$)$^2$ are expected soon
and an experiment for $Q^2 \lesssim 3.4$ (GeV/$c$)$^2$ is in preparation.
These data should improve the accuracy of the neutron charge density
considerably, but data approaching 20 (GeV/$c$)$^2$ will probably be
needed to test scaling in the neutron and in the isospin form factors.

Combining the neutron and proton charge densities, we deduced the up
and down quark radial distributions assuming isospin symmetry and 
neglecting heavier quarks.
This schematic model suggests that the distribution is
slightly broader for up quarks than for down quarks in the proton.
With $\lambda_E=2$ we also observe a statistically significant negative 
density for down quarks near 1 fm that might be attributed to the 
$\bar{d}$ content of the pion cloud.

Although we cannot claim that there is a unique relationship between
form factors and densities, 
expansion of densities in a complete radial bases provides
physically appealing parametrizations of form factor data that are 
applicable over a wide range of $Q^2$.
The use of linear expansions in complete bases minimizes the model
dependence of the fitted form factors and provides more realistic error 
bands in both spatial and momentum representations.
Therefore, even if the identification with static densities is discounted, 
the fitted densities do provide useful parametrizations of the form factors 
nonetheless.  
The choice $\lambda_E=\lambda_M=2$ automatically satisfies pQCD scaling
and provides a natural means for extrapolating form factors to higher
$Q^2$ for the purpose of planning future experiments.

\acknowledgements
We thank O. Gayou for providing a table of $G_{Ep}$ data and X. Ji,
C. Perdrisat, R. Madey, J. Friar, and R. Rosenfelder for useful discussions.
The support of the U.S. National Science Foundation under grant PHY-9971819 
is gratefully acknowledged.


\newpage

\clearpage
\newpage

\begin{table}
\caption{Data selection summary.
\label{table:data}}
\begin{tabular}{lllll}
quantity & reaction & $Q^2$ (GeV/$c$)$^2$ & Ref. & method \\ \hline
$G_{Mp}$ & $p(e,e^\prime)$ & 0.017 -- 0.15 & \cite{Hohler76} & Rosenbluth \\
         &                 & 0.16 -- 31.2 & \cite{Brash02} & 
re-analysis using $G_{Ep}/G_{Mp}$ from recoil polarization \\ \hline
$G_{Ep}$ & $p(e,e^\prime)$ & 0.005 -- 0.055 & \cite{Simon80} & Rosenbluth \\
         &                 & 0.13 -- 1.75& \cite{Price71} & Rosenbluth \\
$G_{Ep}/G_{Mp}$ & $p(\vec{e},e^\prime \vec{p})$ &  0.37 -- 5.54  &
\cite{MKJones00,Gayou02,Pospischil01,Milbrath99} & 
using $G_{Mp}$ from Ref. \cite{Brash02} \\ \hline
$G_{Mn}$ & $d(e,e^\prime n)$ & 0.11 -- 0.26 & \cite{Markowitz93} & 
absolute, efficiency from $d(\gamma,pn)$ \\
& $d(e,e^\prime N)$ & 0.12 -- 0.61 & \cite{Bruins95} & 
ratio method, efficiency from $p(\gamma, \pi^+)n$ \\
& $d(e,e^\prime N)$ & 0.095, 0.126 & \cite{Anklin94} & 
ratio method, efficiency from elastic $p(n,p)n$ \\
& $d(e,e^\prime N)$ & 0.24 -- 0.78 & \cite{Anklin98} & 
ratio method, efficiency from elastic $p(n,p)n$ \\
& & 0.07 -- 0.89 & \cite{Kubon02} & 
ratio method, efficiency from elastic $p(n,p)n$ \\
& $d(e,e^\prime)$ & 1.75 -- 4.0 & \cite{Lung93} & quasielastic \\
& $d(e,e^\prime)$ & 2.5 -- 10.0 & \cite{Rock82} & quasielastic \\ 
& $^3\vec{He}(\vec{e},e^\prime)$ & 0.1, 0.2 & \cite{Xu00} &
Fadeev analysis based upon Refs.\ \cite{Golak95,Golak01} \\ \hline
$G_{En}$ & $d(\vec{e},e^\prime \vec{n})$ & 0.26 & \cite{Eden94b} & PWIA \\
  & $d(\vec{e},e^\prime \vec{n})$ & 0.15, 0.34 & \cite{Ostrick99} 
 & FSI analysis by Ref.\ \cite{Herberg99} \\
 & $\vec{d}(\vec{e},e^\prime n)$ & 0.2 & \cite{Passchier99} & 
FSI from Arenh\"ovel {\it et al}.\ \cite{Arenhovel92} \\
 & $\vec{d}(\vec{e},e^\prime n)$ & 0.5 & \cite{Zhu01} & 
FSI from Arenh\"ovel {\it et al}.\ \cite{Arenhovel92} \\
& $^3\vec{{\rm He}}(\vec{e},e^\prime n)$ & 0.4 & 
\cite{Becker99} & Fadeev analysis by Ref. \cite{Golak01} \\
 & $^3\vec{{\rm He}}(\vec{e},e^\prime n)$ & 0.67 & \cite{Rohe99} & PWIA \\
& $t_{20}$,$T_{20}$ & 0.008 -- 1.64 & \cite{Schiavilla01} & 
extracted from deuteron quadrupole form factor \\ \hline
$\langle r^2_n \rangle$ & $e(n,n)$ & 0 & \cite{Kopecky97} &
thermal neutron transmission in liquid $^{208}$Pb
\end{tabular}
\end{table}

\begin{table}
\caption{Moments, radii, and $\chi^2$ per point for the four fitted densities. 
\label{table:moments}}
\begin{tabular}{lllllllll}
quantity & model & $\lambda$ & $\chi^2/N$ & $M_0$ & $M_2$  (fm$^2$)& 
$\bar{R}$  (fm) & $\xi$  (fm)\\ \hline
$G_{Ep}$ & LGE & 0 & 0.67 & $1.003 \pm 0.001$ & $0.776 \pm 0.019$ & $0.879 \pm 0.011$ & $0.879 \pm 0.011$ \\
         &     & 1 & 0.69 & $1.003 \pm 0.002$ & $0.712 \pm 0.020$ & $0.843 \pm 0.012$ & $0.881 \pm 0.012$ \\
         &     & 2 & 0.71 & $1.003 \pm 0.002$ & $0.649 \pm 0.021$ & $0.804 \pm 0.013$ & $0.883 \pm 0.014$ \\
$G_{Ep}$ & FBE & 0 & 0.69 & $1.003 \pm 0.001$ & $0.776 \pm 0.020$ & $0.880 \pm 0.011$ & $0.880 \pm 0.012$ \\
         &     & 1 & 0.70 & $1.003 \pm 0.002$ & $0.713 \pm 0.024$ & $0.843 \pm 0.014$ & $0.882 \pm 0.015$ \\
         &     & 2 & 0.73 & $1.003 \pm 0.002$ & $0.651 \pm 0.030$ & $0.806 \pm 0.019$ & $0.884 \pm 0.020$ \\ \hline
$G_{Mp}$ & LGE & 0 & 0.71 & $1.012 \pm 0.008$ & $0.729 \pm 0.038$ & $0.849 \pm 0.023$ & $0.849 \pm 0.025$ \\
         &     & 1 & 0.71 & $1.012 \pm 0.008$ & $0.659 \pm 0.040$ & $0.807 \pm 0.024$ & $0.847 \pm 0.027$ \\
         &     & 2 & 0.73 & $1.013 \pm 0.008$ & $0.599 \pm 0.037$ & $0.769 \pm 0.024$ & $0.851 \pm 0.026$ \\
$G_{Mp}$ & FBE & 0 & 0.71 & $1.012 \pm 0.008$ & $0.729 \pm 0.038$ & $0.849 \pm 0.023$ & $0.849 \pm 0.025$ \\
         &     & 1 & 0.71 & $1.012 \pm 0.008$ & $0.661 \pm 0.040$ & $0.808 \pm 0.025$ & $0.848 \pm 0.027$ \\
         &     & 2 & 0.75 & $1.015 \pm 0.009$ & $0.615 \pm 0.073$ & $0.778 \pm 0.046$ & $0.859 \pm 0.050$ \\ \hline
$G_{Mn}$ & LGE & 0 & 2.69 & $1.016 \pm 0.025$ & $0.839 \pm 0.099$ & $0.909 \pm 0.055$ & $0.909 \pm 0.058$ \\
         &     & 1 & 2.69 & $1.027 \pm 0.028$ & $0.823 \pm 0.116$ & $0.895 \pm 0.064$ & $0.931 \pm 0.067$ \\
         &     & 2 & 2.71 & $1.023 \pm 0.027$ & $0.736 \pm 0.107$ & $0.848 \pm 0.063$ & $0.922 \pm 0.065$ \\
$G_{Mn}$ & FBE & 0 & 2.69 & $1.017 \pm 0.026$ & $0.843 \pm 0.101$ & $0.910 \pm 0.056$ & $0.910 \pm 0.059$ \\
         &     & 1 & 2.70 & $1.028 \pm 0.029$ & $0.829 \pm 0.120$ & $0.898 \pm 0.066$ & $0.934 \pm 0.069$ \\
         &     & 2 & 2.75 & $1.026 \pm 0.028$ & $0.752 \pm 0.123$ & $0.856 \pm 0.074$ & $0.930 \pm 0.076$ \\ \hline
$G_{En}$ & LGE & 0 & 0.52 & & $-0.115 \pm 0.003$ \\
         &     & 1 & 0.55 & & $-0.115 \pm 0.003$ \\
         &     & 2 & 0.57 & & $-0.115 \pm 0.003$ \\
$G_{En}$ & FBE & 0 & 0.55 & & $-0.115 \pm 0.003$ \\
         &     & 1 & 0.58 & & $-0.115 \pm 0.003$ \\
         &     & 2 & 0.56 & & $-0.114 \pm 0.003$ \\
\end{tabular}
\end{table}

\end{document}